\documentclass[journal]{IEEEtran}

\usepackage{lineno}
\usepackage{microtype}
\usepackage{graphicx}
\usepackage{subfigure}
\usepackage{booktabs} 
\usepackage{hyperref}
\usepackage{amssymb,amsmath,amsthm,amsfonts,bm}
\usepackage{mathtools}
\usepackage{physics} 
\usepackage{algorithmic} 
\usepackage{algorithm}
\usepackage{adjustbox} 
  
\usepackage{etoolbox}
\makeatletter
\patchcmd{\@makecaption}
  {\scshape}
  {}
  {} 
  {}
\makeatother

\newtheorem{theorem}{Theorem}
\newtheorem{lemma}{Lemma}

\newtheorem{prop}{Proposition}
\newtheorem{definition}{Definition}
\newtheorem{corollary}{Corollary}
\newtheorem{remark}{Remark}

\newcommand{\Tra}{\mathop{\mathrm{Tr}}}
\newcommand{\supp}{\mathop{\mathrm{sup}}}

\newcommand{\stkout}[1]{\ifmmode\text{\sout{\ensuremath{#1}}}\else\sout{#1}\fi}

\usepackage{color}

\begin{document}
 
\title{Quantum Differentially Private Sparse Regression Learning}

\author{Yuxuan Du, Min-Hsiu Hsieh \IEEEmembership{Senior Member,~IEEE}, Tongliang Liu \IEEEmembership{Senior Member,~IEEE}, Shan You \IEEEmembership{Member,~IEEE}, and Dacheng Tao \IEEEmembership{Fellow,~IEEE}
\thanks{
Y.~Du is with School of Computer Science, Faculty of Engineering, The University of Sydney and JD Explore Academy, Beijing 101111, China (e-mail: duyuxuan123@gmail.com). M.~Hsieh is with the Hon Hai Quantum Computing Research Center, Taipei 114, Taiwan. T.~Liu is with the Centre for Quantum Software and Information, Faculty of Engineering and Information Technology, University of Technology Sydney,  Sydney 2006, Australia. S.~You is with the SenseTime, Beijing 100080, China. D.~Tao is with School of Computer Science, Faculty of Engineering, The University of Sydney and JD Explore Academy, Beijing 101111, China.
}
\thanks{M.~Hsieh and D.~Tao are corresponding authors.}
\thanks{Copyright (c) 2017 IEEE. Personal use of this material is permitted. However, permission to use this material for any other purposes must be obtained from the IEEE by sending a request to pubs-permissions@ieee.org.}
}

\markboth{Journal of \LaTeX\ Class Files,~Vol.~14, No.~8, August~2015}
{Shell \MakeLowercase{\textit{et al.}}: Bare Demo of IEEEtran.cls for IEEE Journals}

\maketitle

\begin{abstract}
The eligibility of various advanced quantum algorithms will be questioned if they can not guarantee privacy. To fill this knowledge gap, here we devise an efficient quantum differentially private (QDP) Lasso estimator to solve sparse regression tasks. Concretely, given $N$ $d$-dimensional data points with $N\ll d$, we first prove that the optimal classical and quantum non-private Lasso requires $\Omega(N+d)$ and $\Omega(\sqrt{N}+\sqrt{d})$ runtime, respectively. We next prove that the runtime cost of QDP Lasso is  \textit{dimension independent}, i.e., $O(N^{5/2})$, which implies that the QDP Lasso  can be faster than both the optimal classical and quantum non-private Lasso. Last, we exhibit that the QDP Lasso attains a near-optimal utility bound $\tilde{O}(N^{-2/3})$ with privacy guarantees and discuss the chance to realize it on near-term quantum chips with advantages.
\end{abstract}
 

\IEEEpeerreviewmaketitle

\section{Introduction}
\label{sec:Introduction}
Quantum machine learning (QML), as a burgeoning field in quantum computation, aims to facilitate machine learning tasks with  quantum advantages \cite{biamonte2017quantum}. Numerous theoretical studies have shown that QML algorithms can dramatically reduce the runtime complexity over their classical counterparts, e.g., quantum perceptron \cite{kapoor2016quantum}. Meanwhile, experimental studies have employed near-term quantum devices to accomplish toy-model learning tasks \cite{havlivcek2019supervised,schuld2019quantum,zhu2019training}. Theoretical and experimental results suggest that we are stepping into a new era, in which near-term quantum processors \cite{arute2019quantum,preskill2018quantum} can be employed to benefit practical learning tasks. 

Despite the advance of QML algorithms, the legality of the proposed quantum learning algorithms in many data-sensitive applications will be questioned if they cannot guarantee privacy, caused by legal, financial, or moral reasons. Therefore, to broaden the applicability of QML, it is essential to design quantum private learning algorithms, which can train accurate learning models without exposing precise information of individual training examples such as medical records for patients. However, how to design such quantum algorithms with runtime advantages remains unexplored.

In the classical scenario, differential privacy (DP), which quantitively formalizes  a  rigorous and standard notion of  `privacy', provides  one of the most prominent  solutions towards private learning \cite{dwork2006calibrating}. During the past decade, extensive DP learning  algorithms  have been proposed under  varied practical settings \cite{bassily2014private,chaudhuri2009privacy,jain2012differentially,thakurta2013differentially,wang2019differentially}. Motivated by the success of DP learning and the limited results of quantum private learning, it is highly desired to devise quantum differentially private (QDP) learning algorithms to broaden applications of QML in the data-sensitive areas.

 A fundamental topic in private learning is to devise DP sparse regression learning models \cite{thakurta2013differentially,kasiviswanathan2016efficient}. Let $D=\{\mathbf{X}_i, \bm{y}_i\}_{i=1}^N$ be the given dataset, where $\mathbf{X}_i\in\mathbb{R}^d$ and $ \bm{y}_i\in\mathbb{R}$ are the $i$-th  feature vector and the corresponding target, respectively. Equivalently, we write $D=\{\mathbf{X}\in\mathbb{R}^{N\times d}, \bm{y}\in\mathbb{R}^N\}$. In all practical scenarios, $N\ll d$.  Suppose that $D$ satisfies $\bm{y}=\mathbf{X}\bm{\theta}^*+\bm{\omega}$, where  $\bm{\theta}^*\in\mathbb{R}^d$ with $\|\bm{\theta}^*\|_0=\mathbf{s}$ (i.e., $\mathbf{s}\ll N$) is the underlying sparse parameter to be estimated, and $\bm{\omega}\in\mathbb{R}^d$ is the noise vector. The goal of DP sparse regression is to recover $\bm{\theta}^*$ while satisfying differential privacy. The mainstream learning model to tackle this task is the DP Lasso estimator \cite{talwar2015nearly}, which estimates $\bm{\theta}^*$ by minimizing the loss function $\mathcal{L}(\bm{\theta})$, i.e.,   
\begin{equation} \label{eqn:def_Lasso}
  \arg\min_{\bm{\theta}\in\mathcal{C}} \mathcal{L}(\bm{\theta}):= 
  \frac{1}{2N}  \| \mathbf{X} \bm{\theta} - \bm{y} \|_2^2~,
  \end{equation} 
  where the constraint set $\mathcal{C}=\{\bm{\theta}\in\mathbb{R}^d:\|\bm{\theta}\|_1\leq l_1 \}$ is an $\ell_1$ norm ball that guarantees the sparsity of the estimated result and the differential  privacy should be preserved with respect to any change of the individual pair $(\mathbf{X}_i,\bm{y}_i)$.  Following the conventions \cite{thakurta2013differentially,kasiviswanathan2016efficient,talwar2015nearly,wanglx2019differentially}, we set $l_1=1$, and suppose that  $\|\mathbf{X}\|_\infty\leq 1$ and $\|\bm{y}\|_\infty\leq 1$ throughout the paper for ease of comparison. Note that our results can be easily generalized to $l_1\in\mathbb{R}_+$, $\|\mathbf{X}\|_\infty \leq c_1 $, and $\|\bm{y}\|_{\infty} \leq c_2$, where $c_1, c_2 \in \mathbb{R}_+$ are two positive constants.

 An important utility measure of the private Lasso is the expected excess  risk $R_{\mathcal{L}}$, i.e., 
\begin{equation}
R_{\mathcal{L}}:= \mathbb{E}(\mathcal{L}(\bm{\theta})) - \min_{\bm{\theta}\in\mathcal{C}}\mathcal{L}(\bm{\theta})~,
\end{equation}
where the expectation is taken over the internal randomness of the algorithm for generating $\bm{\theta}$. For example, the studies \cite{kasiviswanathan2016efficient,talwar2015nearly} proposed private Lasso algorithms with a utility bound $\tilde{O}(1/N^{2/3})$. Ref.~\cite{kifer2012private} presented a private algorithm with an $\tilde{O}(1/N)$ utility bound, under the strong convexity and mutual incoherence assumptions. All of these algorithms run in polynomial time in $N$ and $d$.  

\begin{table*}[t]
\centering
\caption{\small{\textbf{The summary of the runtime cost and utility bound of various Lasso estimators on the input size $N$, the feature dimension $d$, and the privacy budget $\epsilon$}. Note that the label `OPT-C', `CDP', `OPT-Q', `QNP', and `QDP' refers to the optimal classical non-private, best-known classical DP, optimal quantum non-private, the proposed quantum non-private, and QDP Lasso estimators,  respectively. For the non-private Lasso, the estimation error $\varsigma$ is set as a constant. }} 
\label{tab:my-table} 
\begin{tabular}{|l|l|l|l|l|l|}
\hline
 & OPT-C   & CDP & OPT-Q & QNP & QDP \\ \hline
Runtime   & $\Omega(N+d)$          &  $O(Nd^2)$              & $\Omega(\sqrt{N}+\sqrt{d})$                   &  $\tilde{O}(\sqrt{Nd})$   &    $ \tilde{O}(N^{5/2}\epsilon^{2})$      \\ \hline
Utility        & ---     & $\tilde{O}\left(\frac{1}{(N\epsilon)^{2/3}}\right)$         & ---              &     --- &       $\tilde{O}\left(\frac{1}{(N\epsilon )^{2/3}}\right)$       \\ \hline
\end{tabular}
\end{table*}  

Different from the classical scenario, how to design QDP Lasso  remains unknown because of the \textbf{disparate priorities} of QML and DP. Specifically, QML algorithms pursue low runtime overhead, while DP learning concerns a good utility. Therefore, the proposed QDP  algorithm should accommodate the following two requirements from each side:
\begin{enumerate}
\item a lower runtime over its classical counterparts;
\item a near-optimal utility bound.
\end{enumerate}

\textbf{Contributions.} 
The main contribution in this study is devising a quantum  \textit{differentially private} Lasso estimator to tackle the private sparse regression learning tasks in Eqn.~(\ref{eqn:def_Lasso}) with a lower runtime complexity than both \textbf{optimal classical and quantum} Lasso estimators and  a \textbf{near-optimal} utility bound. To the best of our knowledge, this is the first quantum private learning algorithm that can  solve practical learning problems with provable advantages.

\textit{\underline{Quantum input/output models}.} To make a fair comparison, the quantum input/output models used in our study are restricted to have an \textbf{almost  identical} setting to the classical case \cite{clarkson2012sublinear}, except that we allow coherent queries to the entries of given inputs. The formal definition of quantum input models used in this study is as follows.
\begin{definition}[Input models]\label{def:input_model}
For a given dataset {$D=\{\mathbf{X}\in\mathbb{R}^{N\times d}, \bm{y}\in\mathbb{R}^N\}$}, the classical  (\textit{quantum}) input oracles $O_{\mathbf{X}}$ and $O_{\bm{y}}$ can   recover  the entry $\mathbf{X}_{ij}$ and the entry $\bm{y}_i$ with $i\in[N]$ and $j\in[d]$ in $O(1)$ time (\textit{in superposition}).
\end{definition} 
Note that `in superposition' means that coherent queries are permitted in the quantum input model, i.e., the algorithm is allowed to  query many locations at the same time  \cite{nielsen2010quantum}. Many quantum algorithms have exploited  this  condition to gain quantum advantages, e.g.,     Ref.~\cite{li2019sublinear} employed the above quantum input/out model to design a sublinear runtime quantum kernel  classifier. 

 Let us further emphasize the importance of the employed quantum input model in Definition \ref{def:input_model}. As questioned by \cite{li2019sublinear}, using  too powerful quantum input models may render the achieved quantum speedup \textit{inconclusive}. Namely, given a strong classical input model, quantum-inspired classical algorithms can collapse many QML algorithms claiming exponential speedups \cite{tang2019quantum}.

Our first theoretical result is analyzing runtime and utility bound of the proposed QDP Lasso estimator. 
\begin{theorem}[Informal, see Theorems \ref{thm:DP_private_lasso_impt} and  \ref{thm:utility_guarantee_impt} 
for the formal description] \label{thm:informal_runtime_private_lasso} 
Given the quantum input models $O_{\mathbf{X}}$ and $O_{\bm{y}}$ in Definition~\ref{def:input_model}, the proposed QDP Lasso is $(\epsilon, \delta)$-differentially private, and outputs $\bm{\theta}^{(T)}$ after $T\sim O((N\epsilon)^{2/3})$ iterations  with overall  $\tilde{O}(N^{5/2}\epsilon^{2})$ runtime and the utility bound 
\[
R_{\mathcal{L}}  = \tilde{O}\left((N\epsilon)^{-2/3}\right).
\] 
\end{theorem}    

To figure out whether the proposed QDP Lasso can attain a runtime speedup, we further explore runtime cost of the optimal classical and quantum non-private Lasso estimators. 
\begin{theorem}[Modified from Lemma \ref{lem:lb_Lasso} and Corollary \ref{coro:Q_Lasso_lw}]\label{thm:informal_runtime_lwb}
	Given the input model formulated in Definition \ref{def:input_model}, the  runtime complexity  of the classical and quantum non-private Lasso estimators is lower bounded by $\Omega(N+d)$ and $\Omega(\sqrt{N}+\sqrt{d})$, respectively.
\end{theorem} 

Theorems \ref{thm:informal_runtime_private_lasso} and  \ref{thm:informal_runtime_lwb} provide the following three key insights.
\begin{itemize}
\item The runtime complexity of QDP Lasso is \textbf{dimension independent}, while the runtime for both the optimal classical and quantum non-private Lasso estimators depends on the feature dimension $d$. In other words, QDP Lasso yields a runtime speedup when $d \geq O(N^{5/2})$ and $d \geq O(N^5)$, respectively.
\item Note that the non-private Lasso requires $O(Nd^2)$ runtime \cite{efron2004least} and assigning the DP property generally involves extra operations. Hence, the best known DP Lasso needs $O(Nd^2)$ runtime, where the QDP Lasso outperforms it when $d>O(N^{3/4})$.
\item  The utility bound of the proposed QDP Lasso is \textbf{near-optimal}, since the optimal utility bound of sparse regression learning has proven to be ${\Omega}(1/(N\log N)^{2/3})$ \cite{talwar2015nearly}.
\end{itemize}
 
We summarize the main conclusions in this work in Table \ref{tab:my-table}.

There are two main technical results that differentiate the QDP Lasso with the classical DP Lasso \cite{talwar2015nearly}. The first one is the quantum generalization of the Frank-Wolfe algorithm for non-private Lasso \cite{frank1956algorithm}, which will be employed as the backbone of the QDP Lasso estimator. We prove that the runtime cost of the proposed algorithm is $\tilde{O}(\sqrt{Nd})$, which achieves a quadratic runtime speedup over the optimal classical Lasso in terms of the feature dimension $d$. Moreover, we conduct error analysis to confirm its stability (see Theorem \ref{thm:runtime_q_lasso} for details). 
 
Our second key technical result is the devised privacy mechanism, which ensures the QDP Lasso estimator to attain privacy guarantee and achieve a runtime speedup over the optimal quantum non-private Lasso. Note that this result \textbf{contradicts} classical DP algorithms, since assigning DP property to non-private algorithms generally requires extra operations. In contrast with the classical scenario, a crucial observation in the QDP Lasso is:\\
\textit{`The probabilistic nature of quantum mechanics enables quantum learning models to efficiently gain the DP property'.}\\
Such an observation indicates that the sampling-based input models \cite{clarkson2012sublinear} may provide substantial runtime advantages for classical DP learning algorithms.

\section{Related work} 
Previous QML literature related to our work can be divided into two groups. The first group is quantum regression algorithms and the second group is quantum differential privacy. Here we separately compare them with our work. 

\textit{\underline{Quantum regression algorithms}.} There are a few proposals aiming to solve quantum regressions tasks without the privacy requirement. A seminal work is \cite{wiebe2012quantum}, which showed that the ordinary least squares fitting problem can be solved with an exponential speedup given the assumption that there exists a  quantum random access memory (QRAM) to encode classical input into quantum states in logarithmic runtime   \cite{giovannetti2008architectures,giovannetti2008quantum}. Under such an assumption,  the quantum linear systems algorithm \cite{harrow2009quantum} can be employed to compute the closed-form expression for the estimated solution with an exponential speedup. Following this pipeline, the subsequent works further improve the runtime complexity bound with respect to the polynomial terms \cite{chakraborty2018power,schuld2016prediction,wang2017quantum}, e.g., rank and condition number,  and tackle variants of the regression tasks, e.g., nonlinear regression and ridge regression  \cite{liu2017fast}. In contrast to solving the closed-form expression, the study \cite{kerenidis2017quantum} tackles the ridge regression tasks by using the gradient descent method, where the runtime complexity achieves the exponential speedup at each iteration under the  QRAM assumption.  We remark that the applicability of these algorithms is highly questionable, since it is still an open question about how to efficiently implement QRAM. Moreover, recent quantum-inspired algorithms adopt the similar assumption of QRAM and dequantize  numerous quantum algorithms with exponential speedups \cite{tang2019quantum,du2019quantum,gilyen2018quantum}.

Several studies developed the hybrid quantum-classical methods to solve regression tasks on near-term quantum devices \cite{mitarai2018quantum,zhang2020prot}. Since there is no theoretical convergence guarantee for these hybrid methods, they are incomparable with our result.   

\textit{\underline{Quantum private learning}.} Several studies have investigated the topic of quantum  private learning  \cite{du2020quantum,zhou2017differential,ying2017quantum,aaronson2019gentle,arunachalam2021private}. In particular, the main contribution of the two studies \cite{aaronson2019gentle,du2020quantum} is to utilize the result of differential privacy to advance quantum tasks, i.e., shadow tomography and defending adversarial attacking. The study \cite{ying2017quantum} proposed a quantum private perceptron, while the privacy metric used in \cite{ying2017quantum} follows the study \cite{agrawal2000privacy}, which is irrelevant to  the notion of differential privacy and is incomparable with our results. The study \cite{zhou2017differential}  developed quantum privacy mechanisms and analyzed their privacy guarantees. However, how to connect QDP with learning algorithms is unexplored.  Recently, the study \cite{arunachalam2020quantum} explored quantum differential privacy from the perspective of learning theory. The study \cite{arunachalam2021private} systematically exploited the connection between differential privacy and algorithms to learn a quantum state. We emphasize that, unlike the above studies, our work aims to develop a private learning algorithm that achieves both the quantum advantage  and the provable utility guarantee. These two factors have not been considered together before.

\section{Preliminaries}\label{sec:preliminaries}

We unify some basic notation throughout the whole paper.  The set  $\{1,2,..., n\}$ is denoted as $[n]$.  Given a matrix $\mathbf{X}\in\mathbb{R}^{N\times d}$ and a vector ${\bm{v}}\in \mathbb{R}^{N}$, the $i$-th row of $\mathbf{X}$ and the $i$-th entry of $\bm{v}$ are represented by $\mathbf{X}_i$ and ${\bm{v}}_i$, respectively. The $\ell_p$ norm of $\mathbf{v}$ ($\mathbf{X}$) is denoted as $\|{\mathbf{v}}\|_p$ ($\|\mathbf{X}\|_p$). The Frobenius norm of $\mathbf{X}$ is defined as $\|\mathbf{X}\|_F=(\sum_{i=1}^n\sum_{j=1}^d|\mathbf{X}_{i,j}|^2)^{1/2}$. The notation $\mathbf{e}_i$ always refers to the $i$-th unit basis vector, e.g., for $\mathbf{e}_i\in\mathbb{R}^3$, $\mathbf{e}_1=[1, 0, 0]$.  The identity matrix of size $D\times D$ is denoted as $\mathbb{I}_D$.  

\subsection{Convex optimization}
We introduce two basic definitions in convex optimization. Refer to \cite{boyd2004convex} for more details. 

\begin{definition}[L-Lipschitz]
A function f is called L-Lipschitz over a set $\mathcal{C}$ if for all $\bm{u},\bm{w}\in \mathcal{C}$, we have
\begin{equation}
|f(\bm{u}) - f(\bm{w})|\leq L\|\bm{u}-\bm{w}\|_2~.
\end{equation}
If $f(\cdot)$ is L-Lipschitz, differentiable, and convex, then
\begin{equation}
\|\nabla f(\bm{u}) \|_2 \leq L~.
\end{equation}
\end{definition}
\begin{definition}[Curvature constant \cite{pmlr-v28-jaggi13}]\label{def:curvature_const}
The curvature constant $C_f$ of a convex and differentiable function $f:\mathbb{R}^d \rightarrow \mathbb{R}$  with respect to a compact domain $\mathcal{C}$ is  
\begin{equation}
C_f := \supp_{\bm{z} = \bm{x}+\gamma(\bm{s}-\bm{x})} \frac{2}{\gamma^2}(f(\bm{z})-f(\bm{x}) - \langle \bm{z}-\bm{x}, \nabla f(\bm{x}) \rangle )~,
\end{equation}   
where $\gamma\in(0,1]$, $\bm{x},\bm{s}\in\mathcal{C}$, and $\langle\cdot, \cdot\rangle$ denotes usual inner product.
\end{definition}

\subsection{Quantum computation}\label{sub:quan_compt}
We present essential background of quantum computation, i.e., quantum states, quantum oracles, and  the complexity measure. We refer to \cite{nielsen2010quantum} for details.
	  
Quantum mechanics works in the Hilbert space $\mathcal{H}$ with $\mathcal{H}\approxeq \mathbb{C}$, where $\mathbb{C}$ represents the complex Euclidean space.  We use {Dirac notation} to denote  quantum states. A \textit{pure quantum state} is defined by a vector $\ket{\cdot}$ (named `ket') with unit length. Specifically,  the state $\ket{\bm{a}}\in\mathbb{C}^d$ is  $\ket{\bm{a}} = \sum_{i=1}^d \bm{a}_i \bm{e}_i =  \sum_{i=1}^d \bm{a}_i \ket{i}$ with $\sum_i |\bm{a}_i|^2=1$, where the  computation basis $\ket{i}$   stands for the unit basis vector $\bm{e}_i\in\mathbb{C}^d$.   The inner product of two quantum states $\ket{\bm{a}}$ and $\ket{\bm{b}}$ is denoted by  $\langle \bm{a}|\bm{b}\rangle$, where $\bra{\bm{a}}$ refers to the conjugate transpose of $\ket{\bm{a}}$. We call a state $\ket{\bm{a}}$ is in \textit{superposition} if the number of  nonzero entries in $\bm{a}$ is larger than one.   Analogous to the `ket' notation,  \textit{density operators} can   be used to describe more general  quantum states.  Given a mixture of $m$ quantum pure states $\ket{\psi_i}\in\mathbb{C}^d$ with the probability $p_i$ and $\sum_{i=1}^m p_i =1$,  the density operator $\rho$   presents the mixed state  $\{p_i, \ket{\psi_i}\}_{i=1}^m$  as $\rho = \sum_{i=1}^m p_i\rho_i$ with $\rho_i =\ket{\psi_i}\bra{\psi_i}\in\mathbb{C}^{d\times d}$ and $\Tra(\rho)=1$.

The basic element in quantum computation is the quantum bit (\textbf{qubit}). A qubit is a two-dimensional quantum state, e.g.,   a qubit can be written as $\ket{\bm{a}} =\bm{a}_1\ket{0}+\bm{a}_2\ket{1}$.   Let $\ket{\bm{b}}$ be an another qubit. The quantum state represented by these two qubits is formulated by the tensor product, i.e., $\ket{\bm{a}}\otimes \ket{\bm{b}}$  as a $4$-dimensional vector. Following conventions, we can also write $\ket{\bm{a}}\otimes \ket{\bm{b}}$  as  $\ket{\bm{a},\bm{b}}$ or $\ket{\bm{a}}\ket{\bm{b}}$. For clearness, we sometimes denote $\ket{\bm{a}}\ket{\bm{b}}$ as $\ket{\bm{a}}_A\ket{\bm{b}}_B$, which means that the qubits $\ket{\bm{a}}_A$ ($\ket{\bm{b}}_B$) is assigned in the quantum register $A$ ($B$). There are two typical quantum operations. The first one is \textit{quantum (logic) gates} that operates on a small number  qubits.  Any quantum gate corresponds to a unitary  transformation and can be stated in the circuit model, e.g., an $n$-qubit quantum gate $U$ with $U^{2^n\times 2^n}$ satisfies $UU^{\dagger}=\mathbb{I}_{2^n}$. The second one is the \textit{quantum measurement}, which aims to extract quantum information such as the computation result into the classical form. Given a density operator $\rho$, the outcome $m$ will be measured with the probability $p_m = \Tra(\mathbf{K}_m\rho\mathbf{K}_m^{\dagger})$ and the post-measurement state will be $\mathbf{K}_m\rho\mathbf{K}_m^{\dagger}/p_m$ with $\sum_b \mathbf{K}_b^{\dagger}\mathbf{K}_b=\mathbb{I}$.   
	
A quantum oracle $O$ can be treated as a `black box', which encapsulates  certain  quantum operations and can be used as the input to another algorithm. The quantum input model $O_{\mathbf{X}}$ refers to a unitary transformation that allows us to access the input  data in superposition, i.e., denote $\mathcal{G}$ as a set of indexes to be queried, we have $O_{\mathbf{X}}(\ket{i,j}\ket{0})=\sqrt{|\mathcal{G}|}^{-1}\sum_{i,j\in\mathcal{G}}\ket{i,j}\ket{\mathbf{X}_{ij}}$   for any $i\in[n]$ and $j\in[d]$. Note that, as with classical computers, the quantum state $\ket{\mathbf{X}_{ij}}$ records the binary string of $\mathbf{X}_{ij}$, i.e., $2\rightarrow\ket{10}$. Similar rules can be applied to $O_{\bm{y}}$. Finally, the  runtime complexity of a quantum algorithm is defined as the number of elementary operations employed in the algorithm. We use $O(\cdot )$ to denote the runtime complexity, or use $\tilde{O}(\cdot )$ that hides the poly-logarithmic factors. We also employ the little $o$ notation, i.e., $f(n)=o(g(n))$, to denote that $f(n)/g(n)\rightarrow 0$.

\subsection{Differential privacy}
We provide the definition of classical and quantum DP. 
\begin{definition}[Differential privacy \cite{dwork2014algorithmic}]\label{def:CDP}
An algorithm $\mathcal{A}$ is $(\epsilon, \delta)$-differential private if for any two neighboring datasets $\mathbf{X}$ and $\mathbf{X}'$ with $\mathbf{X},\mathbf{X}'\in \mathbb{R}^{N\times d}$, and for all measurable sets $\mathcal{O}\subseteq \text{Range}(\mathcal{A})$, the following holds:  
\begin{align}
\Pr(\mathcal{A}(\mathbf{X}) \in \mathcal{O}) \leq e^{\epsilon} \Pr(\mathcal{A}(\mathbf{X}')\in \mathcal{O})+\delta.
\end{align}
Here the neighboring datasets $\mathbf{X}$ and $\mathbf{X}'$  refer that the number of rows in $\mathbf{X}$ that need to be modified (e.g., moved) to get the $\mathbf{X}'$ is one. 
\end{definition}

In this study, we exploit the classical notion of DP to denote the neighboring states $\rho$ and $\sigma$, i.e., $\rho$ and $\sigma$ are prepared by two classical neighboring datasets $X$ and $X'$  given in Definition~\ref{def:CDP}. Suppose that the neighboring datasets $\mathbf{X}$ and $\mathbf{X}'$ in Definition \ref{def:CDP} differ in the $i^*$-th row. Following the Definition \ref{def:input_model}, the explicit form of the neighboring quantum states (unnormalized) yields $\rho= \sum_{ij} \ket{\mathbf{X}_{ij}}\ket{i}\ket{j}$ and $\sigma=  \sum_{ij} \ket{\mathbf{X}_{ij}'}\ket{i}\ket{j}$, where the basis $\ket{\mathbf{X}_{ij}}\ket{i}\ket{j}$ is always the same between $\rho$ and $\sigma$  when $i\neq i^*$.

\section{Quantum non-private Lasso estimator}\label{sec:main_result_nonprivate}
The content in this section can be separated into two parts. First, we devise a  quantum non-private Lasso estimator and prove its runtime complexity. Note that the proposed quantum non-private Lasso estimator will serve as the \textit{backbone} of the QDP Lasso. Second, we analyze the runtime complexity (lower bound) of the optimal classical and quantum Lasso, which will be used to compare with the QDP Lasso. 
 
\subsection{Quantum non-private  Lasso estimator}\label{subsec:quantum_Lasso}

Here we propose a quantum version of the Frank-Wolfe (FW) algorithm \cite{frank1956algorithm} to build a quantum non-private Lasso estimator, which is our \textbf{first main technical contribution}. Through exploiting  the robustness of the FW algorithm, the proposed quantum non-private Lasso attains a quadratic speedup over its classical counterpart (see  Theorem \ref{thm:runtime_q_lasso} and Lemma \ref{lem:lb_Lasso}). Note that the proposed algorithm is a prerequisite to devise the QDP Lasso in Section \ref{sec:main_result}.

\textbf{Classical FW.} Let us first review the FW algorithm   (also known as the conditional gradient method). Recall that the FW algorithm and its variants are representative methods to solve constrained convex optimization tasks and have been broadly used to build non-private  Lasso estimators formulated in Eqn.~(\ref{eqn:def_Lasso}) \cite{pmlr-v28-jaggi13,lacoste2015global,wang2016parallel}.  Furthermore, the study \cite{talwar2015nearly} combines  non-private Lasso estimators with a DP mechanism to build the DP Lasso estimator.

 \begin{algorithm}[h!]  \caption{\small{Frank-Wolfe Algorithm for Lasso \cite{pmlr-v28-jaggi13}}}
   \label{alg:FW}
\begin{algorithmic}[1]
   \STATE {\bfseries Input:} Dataset $D=\{\mathbf{X}\in\mathbb{R}^{N\times d}, \bm{y}\in\mathbb{R}^N\}$ with the quantum input oracles $O_{\mathbf{X}}$ and $O_{\bm{y}}$ in Definition \ref{def:input_model}, the loss $\mathcal{L}\in \mathbb{R}_+$ and the constraint set $\mathcal{C}=\{\bm{\theta}\in\mathbb{R}^d:\|\bm{\theta}\|_1\leq 1 \}$ in Eqn.~(\ref{eqn:def_Lasso}), and the total number of iterations $T\in \mathbb{N}_+$; 
   
   \STATE Randomly choose  $\bm{\theta}^{(1)}\in\mathcal{C}$ with one nonzero entry;
       \FOR{$t=1$ {\bfseries to} $T -1 $}
    \STATE  $\forall s  \in [2d]$,  ${\bm{\alpha}}_s^{(t)}\leftarrow \langle \hat{\bm{e}}_s, \nabla \mathcal{L} (\bm{\theta}^{(t)}) \rangle$;
   \STATE Compute $k^{(t)} = \arg\min_{s \in [2d]} \bm{\alpha}_{s}^{(t)}$ and obtain $\hat{\bm{e}}_{k^{(t)}}$;
    \STATE $\bm{\theta}^{(t+1)}\leftarrow (1-\mu_t)\bm{\theta}^{(t)} + \mu_t \hat{\bm{e}}_{k^{(t)}}$, where $\mu_t = \frac{2}{t+2}$; 
   \ENDFOR 
\STATE {\bfseries Output:} $  \bm{\theta}^{(T)}$   
\end{algorithmic}
\end{algorithm}

\begin{figure}[h!] 
\centering
\includegraphics[width = 0.48\textwidth]{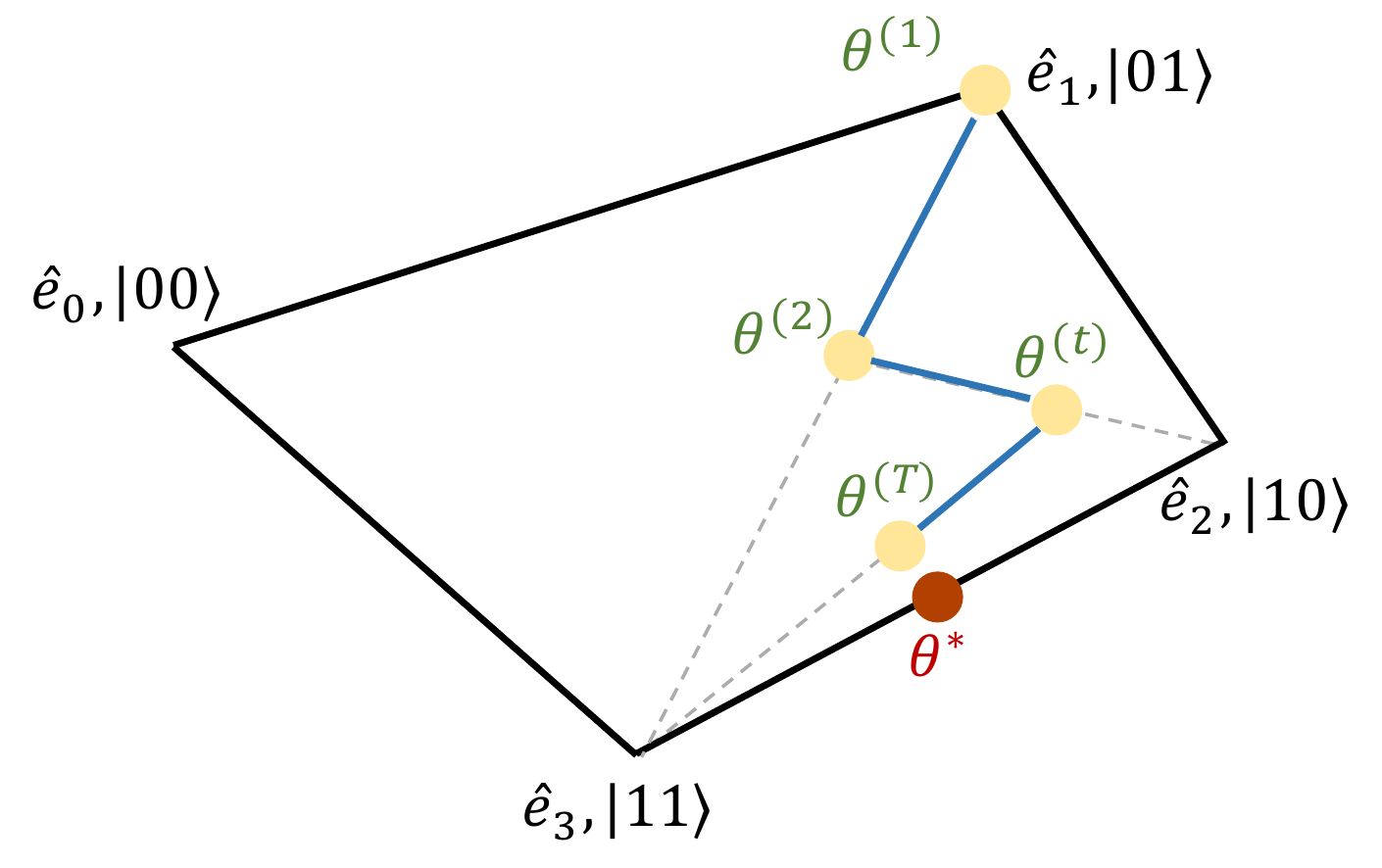}
\caption{{The mechanism of the FW algorithm. The red circle refers to the optimal result $\bm{\theta}^*$. The yellow circle refers to the trainable parameters $\bm{\theta}^{(t)}$ with $t\in[T]$. The initial parameter $\bm{\theta}^{(1)}$ is arbitrarily selected from the vertices set $\mathcal{S}$, e.g., $\hat{\bm{e}}_1$. The grey dash line represents the location of the minimizer $\hat{\bm{e}}_{k^{(t)}}$ at the $t$-th iteration in Eqn.~(\ref{eqn:proj_FW}) (Line 4-5 in Alg.~\ref{alg:FW}). The blue solid line denotes the updating rule from $\bm{\theta}^{(t-1)}$ to $\bm{\theta}^{(t)}$ (Line 6 in Alg.~\ref{alg:FW}). }}
	\label{fig:FW}
\end{figure}

 The implementation of the FW method for Lasso is summarized in  Alg.~\ref{alg:FW}. In detail, FW seeks the target solution $\bm{\theta}^*=\arg\min_{\bm{\theta}\in\mathcal{C}}\mathcal{L}(\bm{\theta})$ in Eqn.~(\ref{eqn:def_Lasso}) through an iteratively updating manner (Line 3-7).  Since the constraint domain $\mathcal{C}$ is an $\ell_1$ norm ball, the optimization can be done by checking each vertex $\hat{\bm{e}}_s$ of the polytope $\mathcal{C}$, where the vertices set is denoted by $\mathcal{S} = \{\hat{\bm{e}}_s\}_{s=1}^{2d}$ so that $\hat{\bm{e}}_s= \bm{e}_{s}$ for $1<s\leq d$ and $\hat{\bm{e}}_s=-{\bm{e}}_{s-d}$ for $d<s\leq 2d$. In other words, the vertices set $\mathcal{S}$  contains $2d$ unit basis vectors $\{\pm  {\bm{e}}_s\}_{s=1}^d$. Figure \ref{fig:FW} illustrates the geometric intuition of the FW algorithm. At the $t$-th iteration, the FW algorithm moves $\bm{\theta}^{(t)}$ to the minimizer  of a linear function (a vertex in the set $\mathcal{S}$), i.e.,
\begin{equation}\label{eqn:proj_FW}
\hat{\bm{e}}_{k^{(t)}}:=  \arg \min_{\hat{\bm{e}}_s\in\mathcal{S}}\langle \hat{\bm{e}}_s, \nabla\mathcal{L}(\bm{\theta}^{(t)}) \rangle~.
\end{equation}
Denote $ \langle \hat{\bm{e}}_s, \nabla \mathcal{L} (\bm{\theta}^{(t)}) \rangle:=\bm{\alpha}_s^{(t)} $ (in Line 4). The explicit form of $\bm{\alpha}_s^{(t)}$  for $s\leq d$  yields   
\begin{align}\label{eqn:sec_state_prep_01}
\bm{\alpha}_s^{(t)} 
 := 
  - \frac{1}{{N}} \sum_{i=1}^n \mathbf{X}_{is} \left(\bm{y}_i    -\langle  \mathbf{X}_i, \bm{\theta}^{(t)} \rangle\right)~.    
\end{align}
For $d<s\leq 2d$, we have $\bm{\alpha}_s^{(t)} =-\bm{\alpha}_{s-d}^{(t)}$. Locating the minimizer $\hat{\bm{e}}_{k^{(t)}}$ is accomplished in Line 5 of Alg.~\ref{alg:FW}.  Note that the updating rule in the FW algorithm is a linear combination of $T$ vertexes (i.e., $\{ \hat{\bm{e}}_{k^{(t)}} \}_{t=1}^T$), which implies that $\bm{\theta}^{(T)}$ is sparse with $\|\bm{\theta}^{(T)}\|_0\leq T$.  

\textbf{Quantum non-private Lasso.} We now introduce the quantum non-private Lasso, where its implementation is summarized in Alg.~\ref{alg:Q_lasso}. In particular, there are two key steps that differ with the classical FW algorithm (Alg.~\ref{alg:FW}); namely, the construction of the oracle $O_{\bm{\alpha}^{(t)}}$ to replace the computation of $\bm{\alpha}^{(t)}$,  and the employment of the quantum minimum finding algorithm \cite{durr1996quantum} to find $\hat{\bm{e}}_{k^{(t)}}$.  These two steps enable the quantum Lasso estimator to quadratically reduce its runtime complexity  to find  $\hat{\bm{e}}_{k^{(t)}}$ for any $t\in[T]$. 

\begin{algorithm}[h!]
 \caption{\small{Quantum Lasso estimator}}
\label{alg:Q_lasso}
\begin{algorithmic}[1]
      \STATE {\bfseries Input:}   Dataset $D=\{\mathbf{X}\in\mathbb{R}^{N\times d}, \bm{y}\in\mathbb{R}^N\}$ with the quantum input oracles $O_{\mathbf{X}}$ and $O_{\bm{y}}$ in Definition \ref{def:input_model}, the loss $\mathcal{L}\in \mathbb{R}_+$ and the constraint set $\mathcal{C}=\{\bm{\theta}\in\mathbb{R}^d:\|\bm{\theta}\|_1\leq 1 \}$ in Eqn.~(\ref{eqn:def_Lasso}), and the total number of iterations $T\in \mathbb{N}_+$;
   
   \STATE Randomly choose  $\bm{\theta}^{(1)}\in\mathcal{C}$ with one nonzero entry;
      \FOR{$t=1$ to $T-1$}
   \STATE Implement the oracle $O_{\bm{\alpha}^{(t)}}:\ket{s}\ket{0}\rightarrow \ket{s}\ket{\bm{\alpha}^{(t)}_s}$ in Lemma \ref{thm:Q_state_prep};
    \STATE  Compute $k^{(t)} = \arg\min_{s \in [2d]} \bm{\alpha}_{s}^{(t)}$ using the quantum minimum finding algorithm in Corollary  \ref{coro:DH_alg_minimum} and the oracle $O_{\bm{\alpha}^{(t)}}$;      \STATE $\bm{\theta}^{(t+1)}\leftarrow (1-\mu_t)\bm{\theta}^{(t)} + \mu_t \hat{\bm{e}}_{k^{(t)}}$, where $\mu_t = \frac{2}{t+2}$; 
    \ENDFOR
     \STATE {\bfseries Output:} $\bm{\theta}^{(T)}$ 
\end{algorithmic}
\end{algorithm}  

We next elaborate on how to implement Lines 4-5 in Alg.~\ref{alg:Q_lasso}.

\textit{\underline{State preparation (Line 4 in Alg.~\ref{alg:Q_lasso})}.} 
This step aims to build the oracle $O_{\bm{\alpha}^{(t)}}$ that  prepares a quantum state related to the classical vector $\bm{\alpha}^{(t)}$ in Eqn.~(\ref{eqn:sec_state_prep_01}) to earn a runtime speedup. 
\begin{lemma}[Oracle $O_{\bm{\alpha}^{(t)}}$]\label{thm:Q_state_prep}
Given access to quantum input models $O_{\mathbf{X}}$ and $O_{\bm{y}}$ in Definition \ref{def:input_model}, the estimated state preparation oracle $O_{\tilde{\bm{\alpha}}^{(t)}}$, which with success probability $1-2b$ and $b\in(0,1)$ prepares a quantum state that  estimates the target state $O_{\bm{\alpha}^{(t)}}\ket{s} \ket{\bm{0}} = \ket{s}\ket{{\bm{\alpha}}^{(t)}_{{s}}}$,   i.e.,
\begin{equation}\label{eqn:thm_state_prep_1}
O_{\tilde{\bm{\alpha}}^{(t)}}\ket{s} \ket{\bm{0}} =  \ket{s}\ket{{\tilde{\bm{\alpha}}}^{(t)}_{{s}}},
\end{equation} 
can be constructed in $\tilde{O}(T^2\sqrt{N}/\varsigma)$ runtime, where $|{\bm{\alpha}}^{(t)}_s - \tilde{\bm{\alpha}}_s^{(t)}|\leq \varsigma$ for any $s\in[2d]$, and the runtime hides a poly-logarithmical term $O(\log(1/b))$. 
\end{lemma}

We defer the construction details of   $O_{\bm{\alpha}^{(t)}}$ and the proof of Lemma \ref{thm:Q_state_prep} in Appendix \ref{appen:sec:state_prep}. Notice that the runtime of calculating the classical vector $\bm{\alpha}^{(t)}$ is at least $O(Nd)$ due to the multiplication of $\mathbf{X}^{\top}\in\mathbb{R}^{d\times N}$ and $(\bm{y} -  \mathbf{X}\bm{\theta}^{(t)})\in\mathbb{R}^{N}$. In contrast, the runtime to prepare the estimated state $\ket{\tilde{\bm{\alpha}}^{(t)}}$ is $\tilde{O}(T^2\sqrt{N}/\varsigma)$ and is independent of {the feature dimension} $d$, celebrated by the coherent property in the quantum input model. Since $T,N\ll d$ in most practical scenarios, this result indicates the efficacy to prepare the state $\ket{\tilde{\bm{\alpha}}^{(t)}}$ instead of directly computing classical form $\bm{\alpha}^{(t)}$, and enables the quantum Lasso to earn a runtime speedup  over the classical Lasso.
 
 \textit{\underline{Find $\hat{\bm{e}}_{k^{(t)}}$ (Line 5 in Alg.~\ref{alg:Q_lasso})} .}  Given access to the oracle $O_{\bm{\alpha}^{(t)}}$, we can directly employ the quantum minimum finding algorithm \cite{durr1996quantum}  to find $k^{(t)}$, or equivalently $\hat{\bm{e}}_{k^{(t)}}$.  We summarize the runtime complexity to find  $\hat{\bm{e}}_{k^{(t)}}$ below.

 \begin{corollary}\label{coro:DH_alg_minimum}
Suppose that the state preparation oracle $O_{\bm{\alpha}^{(t)}}$ in Lemma \ref{thm:Q_state_prep} can be implemented in $T_{\bm{\alpha}}$ runtime. With success probability at least $1/2$, the classical output $\hat{\bm{e}}_{k^{(t)}}$ can be obtained in $\tilde{O}(T_{\bm{\alpha}}\sqrt{d})$ runtime. The success probability can be boosted to $1-1/2^c$ by repeating the quantum minimum finding algorithm $c$ times.  
 \end{corollary}
 
\begin{proof}[Proof of Corollary \ref{coro:DH_alg_minimum}]
Let us first recap a crucial technique used in the quantum non-private Lasso estimator, i.e., the quantum minimum finding algorithm (D{\"u}rr-H{\o}yer's algorithm) \cite{durr1996quantum}.  Recall the quantum minimum finding algorithm \cite{durr1996quantum}. Given an unordered list $\{f(i)\}_{i=1}^{2d}$ with $2d$ items, the goal of the minimum finding algorithm is to find an index $k^*$, i.e., 
\begin{equation}\label{eqn:minimum_find_def}
	k^*=\arg\min_i f(i)~, ~\forall i\in[2d]~.
\end{equation}	 
The theoretical result of  quantum minimum finding algorithm is as follows.
\begin{lemma}[Quantum minimum finding algorithm, 	\cite{durr1996quantum}]\label{lem:durr_qmf_alg}
	The quantum minimum algorithm finds the index $k^*$ defined in Eqn.~(\ref{eqn:minimum_find_def}) with the probability at least $1/2$. The corresponding runtime complexity is $22.5\sqrt{2d}+1.4\log_2^2 (2d)$. 
\end{lemma}  

 \begin{algorithm}[h]
 \caption{\small{Quantum minimum finding algorithm (D{\"u}rr-H{\o}yer’s algorithm)} reformulated}
\label{alg:Q_min_find_Durr_simple}
\begin{algorithmic}[1]
 \STATE {\bfseries Input:}  $\hat{O}$; 
   \STATE Set $T=22.5\sqrt{2d}+1.4\log_2^2 (2d)$, $k\sim [2d]$, and $T'=0$;
      \WHILE{$T'\leq T$	}
      \STATE Initialize the state ($\frac{1}{\sqrt{2d}}\sum_{i=1}^{2d}\ket{i}_{R_1})\otimes\ket{k}_{R_2}$;
      \STATE Use the comparator oracle $O_{comp}$ to mark every item $i$ for which $f(i)\leq f(k)$ with runtime $\hat{T}_1$;
 \STATE Apply Grover search algorithm  to increase the probability of the marked  items with runtime $\hat{T}_2$, and then observe the first register $R_1$ and let $k'$ be the outcome;
 	\STATE $T'\leftarrow T'+\hat{T}_1+\hat{T}_2$;
	 \STATE If $f(k')\leq f(k)$, then set $k\leftarrow k'$;
    \ENDWHILE
    \STATE {\bfseries Output:} $k$ 
\end{algorithmic}
\end{algorithm} 

We follow Ref.~\cite{durr1996quantum} to explain the implementation details of the quantum minimum finding algorithm, summarized in Alg.~\ref{alg:Q_min_find_Durr_simple}, and refer the interested readers to Ref.~\cite{quek2020robust} for the detailed explanation. First, the input of the algorithm is a quantum oracle $\hat{O}$, i.e., $\hat{O}\ket{i}\otimes\ket{0}=\ket{i}\ket{f(i)}$, where $f(i)$ refers to the $i$-th item of the unordered list $\{f(i)\}_{i=1}^{2d}$, denote by $T$ the total runtime. When $T'<T$, the algorithm continuously employs the Grover search to obtain the index $k'$ and compare $f(k)$ and $f(k')$. Once $T'>T$, the quantum minimum finding algorithm outputs $k$ as the prediction of $k^*$.

A central component of the quantum minimum finding algorithm is the comparator oracle $O_{comp}$, which is employed to mark every item with $f(i)\leq f(k)$ for a given $k$. Mathematically, $O_{comp}$ is defined as 
\[
O_{comp}\ket{i}\ket{j}\ket{0}=\ket{i}\ket{j}\ket{g(i,j)},
\]
where $g(i,j)=1$ if $f(i)\leq f(k)$; otherwise, $g(i,j)=0$. Note that $O_{comp}$ can be implemented efficiently by querying the input oracle $\hat{O}$ twice.       

\medskip
We now leverage the result of the  quantum minimum finding algorithm in Alg.~\ref{alg:Q_min_find_Durr_simple} and Lemma \ref{lem:durr_qmf_alg} to prove Corollary \ref{coro:DH_alg_minimum}.

Recall Alg.~\ref{alg:Q_min_find_Durr_simple} and Lemma \ref{lem:durr_qmf_alg}. The runtime of the quantum minimum finding algorithm, i.e., $22.5\sqrt{2d}+1.4\log_2^2 (2d)$,  is dominated by applying Grover search algorithms and  preparing the initialized state, where the first part takes $22.5\sqrt{2d}$ (Line 6 of Alg.~\ref{alg:Q_min_find_Durr_simple}) and the second part takes $1.4\log_2^2 (2d)$ (Line 4 of Alg.~\ref{alg:Q_min_find_Durr_simple}),  respectively \cite{quek2020robust}. However, such a runtime cost is based on the assumption that,  the input oracle can be prepared in $O(1)$ runtime. This is not the case for the quantum Lasso. The construction of the input oracle $O_{\bm{\alpha}^{(t)}}$ to load different entries of $\bm{\alpha}^{(t)}$ takes $T_{\bm{\alpha}}$ runtime. Therefore, the total runtime of the quantum minimum finding algorithm used in the quantum non-private Lasso becomes $\tilde{O}(T_{\bm{\alpha}}\sqrt{2d})$, since the runtime to execute the Grover search algorithms is $\tilde{O}(T_{\bm{\alpha}}\sqrt{2d})$ instead of $O(\sqrt{2d})$. 

Since the success probability of the quantum minimum finding algorithm is $1/2$, with repeatedly querying such an algorithm $c$ times, the probability that none of the $c$ outcomes belong to the minimum result is $1/2^c$. Therefore, with  success probability $1-1/2^c$, there exists at least one target result among $c$ outcomes.    
\end{proof}

In conjunction with Lemma \ref{thm:Q_state_prep} and Corollary \ref{coro:DH_alg_minimum}, we attain a main result of the quantum non-private Lasso estimator.
\begin{theorem} \label{thm:runtime_q_lasso}
Denote $\varsigma$ as the error parameter and $C_f$ as the curvature constant of the loss function $\mathcal{L}$. Given access to $O_{\mathbf{X}}$ and $O_{\bm{y}}$ formulated in Definition  \ref{def:input_model}, with success probability $1-o(1)$, the quantum Lasso as described in  Alg.~\ref{alg:Q_lasso} after $T$ iterations  outputs $\bm{\theta}^{(T)}$ with the utility bound $R_{\mathcal{L}} \leq O(C_f/T+\varsigma)$ in $\tilde{O}(T^3\sqrt{Nd}/\varsigma)$ runtime complexity. 
\end{theorem}

\begin{proof}[Proof of Theorem \ref{thm:runtime_q_lasso}]

The proof of Theorem \ref{thm:runtime_q_lasso} utilizes a technical result of  the Frank-Wolfe algorithm.  A crucial property of the Frank-Wolfe algorithm is its robustness. Specifically, instead of calculating the exact solution $\hat{\bm{e}}_{k^{(t)}}$ as shown in Eqn.~(\ref{eqn:proj_FW}), employing \textit{any} approximated solution $\hat{\bm{e}}_{\tilde{k}^{(t)}}\in\mathcal{S}$ (e.g., obtained by a noisy solver), where $\hat{\bm{e}}_{\tilde{k}^{(t)}}$ is sampled from a certain distribution $P$, to update the learning parameters $\bm{\theta}^{(t)}$ can also promise the convergence of Frank-Wolfe algorithm, as long as  $\hat{\bm{e}}_{\tilde{k}^{(t)}}$  satisfies the following relation,
\begin{equation}\label{eqn:prop_noisy_FW_0_1}
\mathbb{E}_{\hat{\bm{e}}_{\tilde{k}^{(t)}}\sim P}\left[\langle \hat{\bm{e}}_{\tilde{k}^{(t)}}, \nabla \mathcal{L}(\bm{\theta}^{(t)}) \rangle \right] \leq \min_{\hat{\bm{e}}_s\in\mathcal{S}}  \langle \hat{\bm{e}}_s, \nabla \mathcal{L}(\bm{\theta}^{(t)}) \rangle  + \frac{1}{2}\vartheta\mu_t C_f ~,
\end{equation} 
where $C_f$ is the curvature constant  formulated in Definition \ref{def:curvature_const}, $\mu_t$ is the learning rate, and $\frac{1}{2}\vartheta\mu_t C_f$ refers to the additive approximation quality in the step $t$ with $\vartheta\geq 0$ being an arbitrary fixed error parameter \cite{pmlr-v28-jaggi13}. The following proposition quantifies the convergence rate of Frank-Wolfe algorithm. 

\begin{prop}[Theorem 1, \cite{pmlr-v28-jaggi13}]\label{prop:noisy_FW}
Let $\{\hat{\bm{e}}_{\tilde{k}^{(1)}},...,\hat{\bm{e}}_{\tilde{k}^{(T)}}\}$ be a sequence of vectors from $\mathcal{S}$ with $\bm{\theta}^{(t+1)}=(1-\mu_t)\bm{\theta}^{(t)}+\mu_t \hat{\bm{e}}_{\tilde{k}^{(t)}}$, such that for all $t\in[T]$, Eqn.~(\ref{eqn:prop_noisy_FW_0_1}) is satisfied. Then the result $\bm{\theta}^{(T)}$ satisfies
\begin{equation}\label{eqn:prop_noisy_FW_1}
R_{\mathcal{L}} \leq \frac{2C_f}{T+2}(1+\vartheta)~.
\end{equation} 
\end{prop}
We emphasize that, although the original proof of Proposition \ref{prop:noisy_FW} only takes account of the deterministic case, it can be easily extended to the expectation setting given in Eqn.~(\ref{eqn:prop_noisy_FW_0_1}).  Proposition \ref{prop:noisy_FW} implies that the only difference between the exact (i.e., $\hat{\bm{e}}_{k^{(t)}}=\hat{\bm{e}}_{\tilde{k}^{(t)}}$ and $\vartheta=0$) and approximate scenarios (i.e., $\hat{\bm{e}}_{k^{(t)}}\neq \hat{\bm{e}}_{\tilde{k}^{(t)}}$ and $\vartheta>0$) is that the utility bound of the latter is slightly worse than the former. Moreover, under the exact setting, Lasso achieves the utility bound $O(C_f/T)$.

\medskip 
We are now ready to prove Theorem \ref{thm:runtime_q_lasso}.

\textit{Error analysis and utility bound.} The error of  quantum Lasso comes from the two subroutines, Line 4 and Line 5 of Alg.~\ref{alg:Q_lasso}, respectively. First, the state preparation oracle only generates an approximated state $\ket{s}\ket{\tilde{\bm{\alpha}}^{(t)}_s}$ with success probability $1-2b$ and $b\in[0,1]$, as stated in the proof of Lemma \ref{thm:Q_state_prep}. Second, the quantum minimum finding algorithm can only locate the index that corresponds to the minimum entry of $\tilde{\bm{\alpha}}^{(t)}$  with success probability $1-1/2^c$, as shown in Corollary \ref{coro:DH_alg_minimum}.  

Since the quantum minimum finding algorithm queries the oracle $O_{\bm{\alpha}^{(t)}}$ at most $\sqrt{2d}$ times as illustrated in Alg.~\ref{alg:Q_min_find_Durr_simple}, the probability that the state $\ket{s}\ket{\tilde{\bm{\alpha}}^{(t)}_s}$ can always be successfully prepared in all $\sqrt{2d}$ queries is $(1-2b)^{\sqrt{2d}}$. Overall, the success probability to obtain $\hat{\bm{e}}_{\tilde{k}^{(t)}}$ is $(1-1/2^c)(1-2b)^{\sqrt{2d}}$, where the index $\tilde{k}^{(t)}$ is defined as
\begin{equation}\label{eqn:thm_error_analy_QLasso_0}
	\tilde{k}^{(t)} = \arg\min_{s\in[2d]} \tilde{\bm{\alpha}}^{(t)}_s~.
\end{equation}
Since there are in total $T$ iterations in the quantum Lasso algorithm, the success probability to collect $\{\hat{\bm{e}}_{\tilde{k}^{(t)}}\}_{t=1}^T$ is 
\begin{equation}\label{eqn:thm_error_analy_QLasso}
	\left((1-1/2^c)(1-2b)^{\sqrt{2d}}\right)^T=(1-2b)^T (1-2b)^{\sqrt{2d}T}, 
\end{equation}
where $c=\lceil\log_2(1/2b)\rceil$. Eqn.~(\ref{eqn:thm_error_analy_QLasso}) can be simplified as 
\begin{align}\label{eqn:thm_error_analy_QLasso_1}
	& (1-2b)^{T(1+\sqrt{2d})} \geq 1-1/\kappa =1 - o(1),
\end{align}
where we choose $b=\frac{1}{\kappa T(\sqrt{2d}+1)}$, $\kappa>0$. The inequality uses $(1+x/n)^n\geq 1+x$ for $n>1$ and $|x|\leq n$ ($x$ and $n$ correspond to $-1/\kappa$ and $T(\sqrt{2d}+1)$, respectively).  In other words, with success probability $1-o(1)$, we can collect  $\{\hat{\bm{e}}_{\tilde{k}^{(t)}}\}_{t=1}^T$.

We then analyze of the utility bound of quantum Lasso when the collected basis vectors are $\{\hat{\bm{e}}_{\tilde{k}^{(t)}}\}_{t=1}^T$.  Followed from Lemma  \ref{thm:Q_state_prep} and the definition of $k^{(t)}$  as formulated in Line 5 of Alg.~\ref{alg:Q_lasso}, 
we have 
\begin{eqnarray}\label{eqn:proof-thm2-1}
	& \bm{\alpha}_{\tilde{k}^{(t)}}^{(t)}\leq \tilde{\bm{\alpha}}_{\tilde{k}^{(t)}}^{(t)} + \varsigma  \leq \tilde{\bm{\alpha}}_{k^{(t)}}^{(t)} + \varsigma \leq \bm{\alpha}_{k^{(t)}}^{(t)} + 2\varsigma ~,
\end{eqnarray}
where the first inequality uses $|{\bm{\alpha}}_{\tilde{k}^{(t)}}^{(t)} - \tilde{\bm{\alpha}}_{\tilde{k}^{(t)}}^{(t)}|\leq \varsigma$, the second inequality comes from the fact $\tilde{\bm{\alpha}}_{\tilde{k}^{(t)}}^{(t)} =  \min_{s\in[2d]} \tilde{\bm{\alpha}}_s^{(t)}\leq \tilde{\bm{\alpha}}_{k^{(t)}}^{(t)}$, and the last inequality employs  $|{\bm{\alpha}}_{k^{(t)}}^{(t)} - \tilde{\bm{\alpha}}_{k^{(t)}}^{(t)}|\leq \varsigma$. 
By expanding $\bm{\alpha}_{\tilde{k}^{(t)}}^{(t)}$ and $\bm{\alpha}_{k^{(t)}}^{(t)}$ with their explicit forms, we obtain the following relation, i.e., 
\begin{equation}\label{eqn:proof-thm2-0}
\bm{\alpha}_{\tilde{k}^{(t)}}^{(t)}\leq \bm{\alpha}_{k^{(t)}}^{(t)} + 2\varsigma  \Leftrightarrow \langle \hat{\bm{e}}_{\tilde{k}^{(t)}}, \nabla \mathcal{L}(\bm{\theta}^{(t)}) \rangle \leq \min_{\hat{\bm{e}}_s\in\mathcal{S}} \langle \hat{\bm{e}}_s, \nabla \mathcal{L}(\bm{\theta}^{(t)}) \rangle + 2\varsigma~.
\end{equation}
In conjunction with  Eqn.~(\ref{eqn:prop_noisy_FW_0_1}) and (\ref{eqn:proof-thm2-0}), we can choose $\vartheta=\frac{4\varsigma}{\mu_tC_f}$. Finally, Proposition~\ref{prop:noisy_FW} yields
\begin{equation}
R_{\mathcal{L}}\leq \frac{2C_f}{T+2}(1+\vartheta)\leq \frac{2C_f}{T} + 4\varsigma=O(C_f/T+\varsigma)~,
\end{equation}  
where the second inequality employs $\mu_t=2/(T+2)$ and $\vartheta=\frac{4\varsigma}{\mu_tC_f}$.

\textit{Runtime analysis.} We then analyze the runtime complexity of each iteration, which  can be efficiently obtained from Lemma \ref{thm:Q_state_prep} and Corollary \ref{coro:DH_alg_minimum}. As shown in Lemma \ref{thm:Q_state_prep}, the runtime of using the oracle $O_{\bm{\alpha}^{(t)}}$ to prepare the state $\ket{\bm{\alpha}^{(t)}}$ is $T_{\bm{\alpha}}=\tilde{O}((T^2\sqrt{N})/\varsigma)$. Note that we omit the influence of $b$ in the runtime analysis of quantum  Lasso, since the runtime to prepare $O_{\bm{\alpha}^{(t)}}$ only has a logarithmic dependence in terms of $b$. Following the results in the error analysis, at the $t$-th iteration, by repeatedly querying the quantum minimum finding algorithm $c=\lceil\log_2(1/2b)\rceil$ times, the target basis vectors $\hat{\bm{e}}_{\tilde{k}^{(t)}}$ that $\tilde{k}^{(t)}$ satisfies Eqn.~(\ref{eqn:thm_error_analy_QLasso_0}) can be collected with success probability $1-o(1)$. Therefore, based on the claim of Corollary \ref{coro:DH_alg_minimum},  the runtime to find $\hat{\bm{e}}_{\tilde{k}^{(t)}}$ is $\tilde{O}(cT_{\bm{\alpha}}\sqrt{d})$. The runtime of quantum Lasso with $T$ iterations is therefore $\tilde{O}(cTT_{\bm{\alpha}}\sqrt{d})$. By exploiting the explicit form of  $T_{\bm{\alpha}}=\tilde{O}((T^2\sqrt{N})/\varsigma)$  and  $c=\lceil\log_2(1/2b)\rceil$, the runtime complexity of quantum Lasso is then equal to
\[\tilde{O}(T^3\sqrt{Nd}/\varsigma) ~.\]  
\end{proof}

\subsection{The runtime lower bounds in non-private settings}

We end this section by proving the optimal (lower bound) runtime complexity of classical and quantum non-private Lasso estimators with the input model in Definition \ref{def:input_model}. The central tool toward this goal is the equivalence between Lasso and support vector machine (SVM) \cite{jaggi2014equivalence}. This equivalence enables us to leverage the advanced results of the optimal (quantum) SVM \cite{clarkson2012sublinear,li2019sublinear} to infer the runtime of the optimal (quantum) non-private Lasso. The derived results allow us to evaluate performance of the quantum non-private Lasso and the QDP Lasso. 

The runtime cost of the optimal classical Lasso is as follows. 
 \begin{lemma}\label{lem:lb_Lasso}
 Given access to the input model in Definition \ref{def:input_model}, the  runtime complexity  of the classical non-private Lasso is lower bounded by $\Omega(N+d)$. 
 \end{lemma}
 \begin{proof}[Proof of Lemma \ref{lem:lb_Lasso}]
 Let us first recall the result achieved in \cite{jaggi2014equivalence}. In particular, the dual optimization problem of support vector machine (SVM) and its variants (i.e., kernel SVM, soft-margin SVM variants using $l_2$-loss, and  one-class SVMs)  is of the form
\begin{equation}\label{eqn:2}
	\min_{\bm{\theta}\in\Delta } \|\mathbf{X}\bm{\theta}\|^2~,
\end{equation}
where $\Delta$ is the unit simplex in $\mathbb{R}^d$.  The key observation in Sec.~3 of \cite{jaggi2014equivalence} is that the optimization problem in Eqn.~(\ref{eqn:2}) is equivalent to  Lasso as defined in Eqn.~(\ref{eqn:def_Lasso}), i.e., 
	$\min_{\bm{\theta}\in \mathcal{C}} \|\mathbf{X}\bm{\theta}- \bm{y}\|^2$. In particular, \cite{jaggi2014equivalence} states that any instance of Lasso can be reformulated as a $l_2$-loss soft-margin (or hard-margin) SVM instance with the same optimal solutions. This result guarantees the equivalence between SVM and Lasso even though the relaxations are considered. To be more specific, as addressed in Section 4.1 of \cite{jaggi2014equivalence}, the SVM algorithm proposed in \cite{clarkson2012sublinear} can be directly employed to design a sublinear time algorithm for the Lasso.

Following the above observation, we now employ the result obtained from SVM study to quantify the lower bound runtime complexity of the classical Lasso estimator. In particular, the study \cite{clarkson2012sublinear} proves that, given the input model formulated in Definition \ref{def:input_model}, the optimal runtime for the support vector machine (SVM) is $\Omega(N+d)$. Supported by the equivalence between SVM and Lasso, the optimal runtime for Lasso is lower bounded by $\Omega(N+d)$. 	
\end{proof}

We further quantify the runtime complexity of the optimal quantum non-private Lasso estimator.
\begin{corollary}\label{coro:Q_Lasso_lw}
When the estimation error  $\varsigma=\Theta(1)$ is set as a constant, the runtime complexity of the quantum non-private  Lasso is lower bounded by $\Omega(\sqrt{N}+\sqrt{d})$.
\end{corollary}
\begin{proof}[Proof of Corollary \ref{coro:Q_Lasso_lw}]
The proof of this corollary follows Lemma \ref{lem:lb_Lasso} closely. Given the input model formulated in Definition \ref{def:input_model}, the lower bound of quantum SVM is $\Omega(\sqrt{N}+\sqrt{d})$ \cite{li2019sublinear}. In favor of the equivalence between Lasso and SVM \cite{jaggi2014equivalence}, then the runtime lower bound of the corresponding quantum Lasso is also $\Omega(\sqrt{N}+\sqrt{d})$.
\end{proof}

We now use the above two results to assess the performance of the quantum non-private Lasso, while the analysis of the QDP Lasso will be deferred to the next section. In particular, based on Lemma \ref{lem:lb_Lasso}, the proposed quantum non-private Lasso in Theorem \ref{thm:runtime_q_lasso} achieves a \textit{quadratical runtime speedup} over the optimal classical Lasso in terms of the feature dimension $d$.   Moreover, the runtime complexity of the quantum non-private Lasso is \textit{near-optimal} when $N\ll d$, supported by the results of Corollary \ref{coro:Q_Lasso_lw}.

 \section{Quantum private Lasso estimator}\label{sec:main_result}
To ease understanding, let us first review the general rule of transforming non-private learning algorithms to DP algorithms before elaborating on the proposed QDP Lasso. Concretely,  there are three mainstream strategies to assign DP properties to a classical learning algorithm, i.e.,  injecting randomness to  1) the input datasets; 2) the  optimization information such as gradients; and 3) the employed loss function \cite{chaudhuri2011differentially}.   For example, the DP Lasso \cite{talwar2015nearly} attains the privacy promise by adding noise sampled from Laplacian or Gaussian distributions with certain variances to $\bm{\alpha}^{(t)}$ in Line 4 of Alg.~\ref{alg:FW}. 

Due to the fact that injecting noise into non-private algorithm generally involves extra operations, almost all current classical DP learning algorithms have a worse runtime complexity than their non-private counterparts. However, the probabilistic nature of quantum mechanics motivates us to rethink: \\ \textit{Can we use such  randomness to attain the DP property instead of injecting handcrafted noise?} 

The proposed QDP Lasso provides a positive affirmation towards the above question. As summarized in Alg.~\ref{alg:Q_lasso_impt_pri_me}, the QDP Lasso adopts the following two simple procedures instead of introducing extra cumbersome operations as classical DP algorithms do to impose DP property on the quantum non-private Lasso  in Alg.~\ref{alg:Q_lasso}. First, the quantum minimum finding algorithm in Line 5 of Alg.~\ref{alg:Q_lasso} is abandoned in the QDP Lasso. Second, the searched index $k^{(t)}$ of the QDP Lasso is acquired by sampling from an \textit{engineered state} $\ket{\bm{\alpha}^{(t)}}$ denoted by  $\ket{\mathcal{W}(\bm{\alpha}^{(t)})}$.  

In the rest of this section, we first introduce the quantum subroutine $\mathcal{W}(\bm{\alpha}^{(t)})$ in Line 7 of Alg.~\ref{alg:Q_lasso_impt_pri_me} that prepares the state $\ket{\mathcal{W}(\bm{\alpha}^{(t)})}$. We next analyze runtime and utility of the proposed QDP Lasso.  Last, we elaborate on the necessity of the involved examination procedure Line 3-5 of Alg.~\ref{alg:Q_lasso_impt_pri_me}.

\begin{algorithm}[h!]
 \caption{\small{$(\epsilon,\delta)$-QDP Lasso estimator}}
\label{alg:Q_lasso_impt_pri_me}
\begin{algorithmic}[1]
 \STATE  {\bfseries Input:}  
 Dataset $D=\{\mathbf{X}\in\mathbb{R}^{N\times d}, \bm{y}\in\mathbb{R}^N\}$ with the quantum input oracles $O_{\mathbf{X}}$ and $O_{\bm{y}}$ in Definition \ref{def:input_model}, the loss $\mathcal{L}\in \mathbb{R}_+$ and the constraint set $\mathcal{C}=\{\bm{\theta}\in\mathbb{R}^d:\|\bm{\theta}\|_1\leq 1 \}$ in Eqn.~(\ref{eqn:def_Lasso}), the total number of iterations $T\in\mathbb{N}_+$, the Lipschitz constant $L_1\in \mathbb{R}_+$, the error threshold $\varsigma\in \mathbb{R}_+$, and the differential privacy parameters $(\epsilon, \delta)$; \; 
 \STATE Calculate the the hyper-parameter $\lambda = \sqrt{2T\ln(1/\delta)}\frac{8}{\epsilon N}\in \mathbb{R}_+$ in Eqn.~(\ref{eqn:DP_private_lasso_impt}); \;            
 \STATE  Randomly choose  $\bm{\theta}^{(1)}\in\mathcal{C}$ with one nonzero entry; \;
 \IF{$L_1/\lambda \geq  \ln(1/\varsigma)$}
 \STATE  { Break; \; 
   }
   \ENDIF
  \FOR{$t=1$ to $T-1$} 
   { 
 \STATE  Prepare the state $\ket{\mathcal{W}(\bm{\alpha}^{(t)})}$ in Eqn.~(\ref{eqn:impt_state}) by the quantum subroutine $\mathcal{W}(\bm{\alpha}^{(t)})$ in Lemma \ref{lem:Q_state_prep_impt}; \;
 \STATE    Measure the index register $\ket{s}$ of $\ket{\mathcal{W}(\bm{\alpha}^{(t)})}$ conditioned on seeing the last qubit as 0, and set the received index $k^{(t)}\in[2d]$ as $K^{(t)}=k^{(t)} $; \;
 \STATE    $\bm{\theta}^{(t+1)}\leftarrow (1-\mu_t)\bm{\theta}^{(t)} + \mu_t \hat{\bm{e}}_{K^{(t)}}$, where $\mu_t = \frac{2}{t+2}$; \; 
  }
   \ENDFOR   
 \STATE {\bfseries Output:} $\bm{\theta}^{(T)}$
\end{algorithmic}
\end{algorithm}

The aim of the well-designed subroutine $\mathcal{W}(\bm{\alpha}^{(t)})$ in Line 7 of Alg.~\ref{alg:Q_lasso_impt_pri_me} is to achieve the DP property. Mathematically, $\mathcal{W}(\bm{\alpha}^{(t)})$ prepares a superposition state $  \ket{\mathcal{W}(\bm{\alpha}^{(t)})}$, i.e., 
\begin{equation}\label{eqn:impt_state}
  \ket{\mathcal{W}(\bm{\alpha}^{(t)})}=\sum_{s=1}^{2d}\frac{1}{\sqrt{2d}}\ket{s}\ket{\bm{\alpha}^{(t)}_s}  \left(e^{-\frac{|\bm{\alpha}^{(t)}_s+2L_1|}{2\lambda}} \ket{0}+ \ket{\perp_s}\right), 
   \end{equation} 
   where $\ket{\perp_s} =  \sqrt{  1 - e^{-\frac{|\bm{\alpha}^{(t)}_s+2L_1|}{\lambda}} }    \ket{1}$ refers to the garbage state. Based on the explicit form of $\ket{\mathcal{W}(\bm{\alpha}^{(t)})}$, we obtain the following two observations. On the one side, when the index qubit $\ket{s}$ of the quantum state $\ket{\mathcal{W}(\bm{\alpha}^{(t)})}$ is measured, the index $s\in[2d]$ whose corresponding entry $\bm{\alpha}^{(t)}_{s}$ is  close to $\bm{\alpha}^{(t)}_{k^*}=\min_{s\in [2d]}  \bm{\alpha}^{(t)}_s$ will be chosen as $K^{(t)}$ with high probability. On the other side, due to the non-zero probability amplitude,  all indices $s\in[2d]$ could be selected as the output $K^{(t)}$. As shown in Theorem \ref{thm:DP_private_lasso_impt}, these two properties ensure the QDP Lasso to meet the requirements of differential privacy in Definition \ref{def:CDP}.  
    
\begin{theorem}[Privacy guarantee]\label{thm:DP_private_lasso_impt}
Given the dataset $D=\{\mathbf{X}\in\mathbb{R}^{N\times d}, \bm{y}\in\mathbb{R}^N\}$ in Definition \ref{def:input_model}, the QDP Lasso, as illustrated in Alg.~\ref{alg:Q_lasso_impt_pri_me}, achieves $(\epsilon, \delta)$-differential privacy after $T$ iterations, with setting 
\begin{equation}\label{eqn:DP_private_lasso_impt}
\lambda=\sqrt{2T\ln(1/\delta)}\frac{8}{\epsilon N}~.
\end{equation}
\end{theorem} 
 
 \begin{proof}[Proof of Theorem \ref{thm:DP_private_lasso_impt}]

A technical tool employed to prove Theorem \ref{thm:DP_private_lasso_impt} is the strong composition property in differential privacy. 
\begin{prop}[Strong composition, Theorem 3.3, \cite{dwork2010boosting}]\label{prop:stong_compo_thm} 
For every $\epsilon'>0$, $\delta,\delta'>0$, and $k\in \mathbb{N}$, the class of $(\epsilon', \delta')$-differentially private mechanism is $(\epsilon, k\delta' + \delta)$-differentially private under $k$-fold adaptive composition, i.e., 
\begin{align} 
	\epsilon =  \sqrt{2k\ln(1/\delta)}\epsilon' + k\epsilon'\epsilon_0~,
\end{align} 	
where $\epsilon_0 = e^{\epsilon'} -1$. 
 \end{prop}

We are now ready to prove Theorem \ref{thm:DP_private_lasso_impt}.

\medskip	
Here we first quantify the differential privacy of the QDP Lasso at the $t$-iteration and then employ the strong composition property as formulated in Proposition \ref{prop:stong_compo_thm} to demonstrate the privacy guarantee of the proposed QDP Lasso estimator. 
	
To ease notation, we refer Line 7-8 of Alg.~\ref{alg:Q_lasso_impt_pri_me} as the privacy mechanism  $\mathcal{M}$. At the $t$-th iteration, when the input dataset is $D$, denoted $Z^{(t)}= \sum_{s=1}^{2d}\exp\left(-\frac{\left|\bm{\alpha}^{(t)}_{s}+2 L_1 \right|}{\lambda} \right)$, the index $k^{(t)}\in[2d]$ will be accepted as the output $K^{(t)}=k^{(t)}$ with  probability
	\begin{equation}
	\Pr(\mathcal{M}(D)=k^{(t)})= \frac{1}{Z^{(t)}}\exp\left(\frac{-\left|\bm{\alpha}^{(t)}_{k^{(t)}}+2L_1\right|}{\lambda}\right)~.
	\end{equation}
	When the same rule applies to the neighboring dataset $D'$, denoted $Z^{\prime (t)}= \sum_{s=1}^{2d}\exp\left(-\frac{\left|\bm{\alpha}^{\prime(t)}_{s}+2 L_1 \right|}{\lambda} \right)$, the same index $k$ will be accepted as the output $K^{(t)}=k^{(t)}$ with probability
		\begin{equation}
	\Pr(\mathcal{M}(D')=k^{(t)})= \frac{1}{Z^{\prime (t)}}\exp\left(\frac{-\left|\bm{\alpha}^{\prime(t)}_{k^{(t)}}+2L_1\right|}{\lambda}\right) ~.
	\end{equation}

Based on the differential privacy formulated in Definition \ref{def:CDP}, we now bound the ratio 
\begingroup
\allowdisplaybreaks
\begin{align}
&\frac{\Pr(\mathcal{M}(D)=k^{(t)})}{\Pr(\mathcal{M}(D')=k^{(t)})} \nonumber\\
 = & \frac{	 \exp\left(-\left|\bm{\alpha}^{(t)}_{k^{(t)}}+2L_1\right|/\lambda \right) Z^{\prime (t)}}{ \exp\left(-\left|\bm{\alpha}^{\prime(t)}_{k^{(t)}}+2L_1\right|/\lambda \right) Z^{(t)} } \nonumber\\
 = & \frac{ \exp\left(-\left|\bm{\alpha}^{(t)}_{k^{(t)}}+2L_1\right|/\lambda \right)}{\exp\left(-\left|\bm{\alpha}^{\prime(t)}_{k^{(t)}}+2L_1\right|/\lambda \right)  } \frac{\sum_{s=1}^{2d} \exp\left(-\left|\bm{\alpha}^{\prime(t)}_{s}+2L_1\right|/\lambda \right)}{\sum_{s=1}^{2d} \exp\left(-\left|\bm{\alpha}^{(t)}_{s}+2L_1\right|/\lambda \right)  } \nonumber\\
 \leq &   \exp\left(\frac{-\left|\bm{\alpha}^{(t)}_{k^{(t)}}+2L_1\right| + \left|\bm{\alpha}^{\prime(t)}_{k^{(t)}}+2L_1\right|}{\lambda} \right) \nonumber\\
 & \times \frac{\sum_{s=1}^{2d} \exp\left(-\left|\bm{\alpha}^{(t)}_{s}+2L_1\right|/\lambda \right)\exp\left(4/(\lambda N) \right)}{\sum_{s=1}^{2d} \exp\left(-\left|\bm{\alpha}^{(t)}_{s}+2L_1\right|/\lambda \right)  }    \nonumber\\
 \leq & \exp\left(\frac{\left|\bm{\alpha}^{\prime(t)}_{k^{(t)}} + 2L_1 - \bm{\alpha}^{(t)}_{k^{(t)}}-2L_1 \right|}{\lambda} \right)\exp\left(\frac{4}{\lambda N}\right) \nonumber\\
	\leq & \exp\left(\frac{8}{\lambda N} \right) ~,
\end{align} 
\endgroup
where the first inequality uses  $|\bm{\alpha}^{(t)}_{s}-\bm{\alpha}^{\prime(t)}_{s}|\leq \frac{4}{N}$ as proved below and $|\bm{\alpha}^{\prime (t)}_{s}| \leq L_1$ for $\forall s\in[2d]$, the second inequality comes from $||a|-|b||\leq |a-b|$, and the last inequality employs $|\bm{\alpha}^{(t)}_{s}-\bm{\alpha}^{\prime(t)}_{s}|\leq \frac{4}{N}$ again.  In particular, the difference of $\bm{\alpha}_s^{(t)}$ and $\bm{\alpha}_s^{\prime(t)}$ for each entry $s\in[2d]$ is upper bounded by
\begingroup
\allowdisplaybreaks
\begin{align}\label{eqn:app_thm_priv_adap_0}
& |\bm{\alpha}_s^{(t)}-\bm{\alpha}_s^{\prime(t)}| \nonumber\\
= & \Big | - \frac{1}{{N}} \sum_{i=1}^N \mathbf{X}_{is} \left(\bm{y}_i    -\langle  \mathbf{X}_i, \bm{\theta}^{(t)} \rangle\right)  \nonumber\\ 
& +   \frac{1}{{N}} \sum_{i=1}^N \mathbf{X}'_{is} \left(\bm{y}'_i    -\langle  \mathbf{X}_i^{\prime } , \bm{\theta}^{(t)} \rangle\right) \Big| \nonumber\\
\leq & \frac{1}{N}\left|\mathbf{X}_{ks} \right| \left|  \bm{y}_k    -\langle \mathbf{X}_k, \bm{\theta}^{(t)} \rangle \right| +   \frac{1}{N}\left| \mathbf{X}_{ks}'\right| \left|  \bm{y}_k'    -\langle  \mathbf{X}_k^{\prime } , \bm{\theta}^{(t)} \rangle \right| \nonumber \\
\leq & \frac{1}{N} \left|\mathbf{X}_{ks} \right| (\left|  \bm{y}_k  \right| + \left|\langle \mathbf{X}_k, \bm{\theta}^{(t)} \rangle \right| ) + \frac{1}{N} \left|\mathbf{X}_{ks} \right| (\left|  \bm{y}_k' \right| +\left|\langle \mathbf{X}_k', \bm{\theta}^{(t)} \rangle \right| ) \nonumber\\
\leq & \frac{1}{N} \|\mathbf{X}\|_{F} (1 + \|\mathbf{X}\|_{F} ) + \frac{1}{N} \|\mathbf{X}'\|_{F} (1 +\|\mathbf{X}'\|_{F} ) \nonumber\\
\leq & \frac{4}{N}~.  
\end{align}
\endgroup
The first equality in  Eqn.~(\ref{eqn:app_thm_priv_adap_0}) comes from Eqn.~(\ref{eqn:sec_state_prep_01}). The first inequality employs the triangle inequality and the fact that only one example, said the $k$-th example, in $D$ and $D'$ is varied. The second inequality uses the triangle inequality. The third inequality is guaranteed by the facts that $|\mathbf{X}_{ks}|\leq \|\mathbf{X}\|_F$, $|\mathbf{X}'_{ks}|\leq \|\mathbf{X}'\|_F$, $|\bm{y}_k|\leq \|\bm{y}\|\leq 1$, $|\bm{y}'_k|\leq \|\bm{y}'\|\leq 1$, and $\|\mathbf{X}_k\|\|\bm{\theta}^{(t)}\|\leq \|\mathbf{X}\|_F$ and $\|\mathbf{X}'_k\|\|\bm{\theta}^{(t)}\|\leq \|\mathbf{X}'\|_F$ (due to $\|\bm{\theta}^{(t)}\|\leq 1$, see Eqn.~(\ref{eqn:def_Lasso})).   The last inequality  exploits the results $\|\mathbf{X}\|_F\leq 1$ and $\|\mathbf{X}'\|_F\leq 1$.

To ensure the ratio $\frac{\Pr(\mathcal{M}(D)=k^{(t)})}{\Pr(\mathcal{M}(D')=k^{(t)})}$ is upper bounded by $e^{\epsilon'}$, we have to bound
\begin{align}
	& \exp\left(\frac{8}{\lambda N} \right) \leq e^{\epsilon'} \Rightarrow \epsilon' \geq \frac{8}{\lambda N}~.  
\end{align} 
Then, with employing the strong composition property given in Proposition \ref{prop:stong_compo_thm}, the proposed quantum Lasso estimator achieves  $(\epsilon, \delta)$-privacy after $T$ iterations, where
\begin{equation}
	\epsilon \approx \sqrt{2T\ln(1/\delta)}\frac{8}{\lambda N} ~.
\end{equation}
In other words, with setting \[\lambda=\sqrt{2T\ln(1/\delta)}\frac{8}{\epsilon N}~,\] the proposed quantum private Lasso with weighted sampling mechanism achieves $(\epsilon, \delta)$-differential privacy.
\end{proof}

We remark that the devised \textit{privacy mechanism} used in the QDP Lasso, as \textbf{our second main technical contribution}, amounts to sampling $k^{(t)}$ from the state $\ket{\mathcal{W}(\bm{\alpha}^{(t)})}$ in Line 8 of Alg.~\ref{alg:Q_lasso_impt_pri_me}. Mathematically, at the $t$-th iteration, the quantum subroutine $\mathcal{W}(\bm{\alpha}^{(t)})$ as will be described below is employed to prepare $\ket{\mathcal{W}(\bm{\alpha}^{(t)})}$ in Eqn.~(\ref{eqn:impt_state}). This  privacy mechanism is a crucial element to \textbf{dramatically reduce} runtime of the QDP Lasso to outperform both the optimal classical and quantum non-private Lasso estimators, since it avoids involving extra operations such as noise injection and removes the quantum minimum finding algorithm, which dominates the runtime complexity of the quantum non-private Lasso.

We next exhibit that the subroutine $\mathcal{W}(\bm{\alpha}^{(t)})$ can be efficiently implemented. This result is the precondition to assure the efficacy of the proposed QDP Lasso. Based on Eqn.~(\ref{eqn:impt_state}), we observe that the state $ \ket{\mathcal{W}(\bm{\alpha}^{(t)})}$ can be prepared by  applying the conditional rotation operation \cite{harrow2009quantum} to the state  $\ket{\bm{\alpha}^{(t)}}$ accompanied with an ancillary qubit. In other words, the subroutine $\mathcal{W}(\bm{\alpha}^{(t)})$ can be effectively constructed by employing the oracle $O_{\bm{\alpha}^{(t)}}$ in Theorem \ref{thm:Q_state_prep} with extra $O(1)$ overhead. The following lemma summarizes the runtime cost of implementing $\mathcal{W}(\bm{\alpha}^{(t)})$.   
 \begin{lemma}\label{lem:Q_state_prep_impt}
Following notations used in Lemma \ref{thm:Q_state_prep}, there exists a quantum algorithm which with success probability $1-2b$ and $b\in(0,1)$, prepares a quantum state
\begin{equation}
\ket{\widetilde{\mathcal{W}}(\bm{\alpha}^{(t)})}=	\sum_{s=1}^{2d}\frac{1}{\sqrt{2d}}\ket{s}\ket{\tilde{\bm{\alpha}}^{(t)}_s}\left(e^{-\frac{|\tilde{\bm{\alpha}}^{(t)}_s+2L_1|}{2\lambda}}\ket{0}+ \ket{\tilde{\perp}_s}\right),
\end{equation}
where $\ket{\tilde{\perp}_s}=\sqrt{1 - e^{-\frac{|\tilde{\bm{\alpha}}^{(t)}_s+2L_1|}{\lambda}} } \ket{1}$ in $\tilde{O}(T^2\sqrt{N}/\varsigma)$ runtime. Moreover, the prepared state $\ket{\widetilde{\mathcal{W}}(\bm{\alpha}^{(t)})}$   estimates the state $ \ket{\mathcal{W}(\bm{\alpha}^{(t)})}$ in Eqn.~(\ref{eqn:impt_state}) with $|{\bm{\alpha}}^{(t)}_s - \tilde{\bm{\alpha}}_s^{(t)}|\leq \varsigma$ , $\varsigma\leq L_1$, and $\tilde{\bm{\alpha}}_s^{(t)}=-\tilde{\bm{\alpha}}_{s+d}^{(t)}$ for any $s\in[2d]$. 
\end{lemma}
 \begin{proof}[Proof of Lemma \ref{lem:Q_state_prep_impt}]
Let us first recall the construction of the quantum subroutine $\mathcal{W}(\bm{\alpha}^{(t)})$, as the backbone of Alg.~\ref{alg:Q_lasso_impt_pri_me}. Specifically, we first exploit the oracle $O_{\bm{\alpha}^{(t)}}$ in Lemma \ref{thm:Q_state_prep} and quantum arithmetic operations such as addition and multiplication to prepare a uniform superposition state 
\begin{equation}
\ket{\varphi(\bm{\alpha}^{(t)})}=\frac{1}{\sqrt{2d}}\sum_{s=1}^{2d}\ket{s}\ket{\text{exp}(-|\tilde{\bm{\alpha}}^{(t)}_s+2L_1|/\lambda)}~. 	
\end{equation}
We then introduce an ancillary qubit to the state $\ket{\varphi(\bm{\alpha}^{(t)})}$ and rotate the state conditioned on $\text{exp}(-|\tilde{\bm{\alpha}}^{(t)}_s+2L_1|/\lambda)/Z^{(t)}$ to obtain the target state $\ket{\mathcal{W}(\bm{\alpha}^{(t)})}$.

Here we first detail the implementation of the above two steps and then analyze their runtime cost step by step.

The preparation of the state $\ket{\varphi(\bm{\alpha}^{(t)})}$ can be accomplished with an extra $O(1)$ runtime complexity if we have access to the oracle $O_{\hat{\bm{\alpha}}^{(t)}}$ defined in Eqn.~(\ref{eqn:app_a_state_prep_0}). In particular, we first append an ancillary qubit and apply $H$ gate to create the state
\begin{equation}\label{eqn:impt_a_state_prep_0-0}
H\otimes O_{\hat{\bm{\alpha}}^{(t)}} |0\rangle\otimes \ket{\bf{0}} \to \frac{1}{\sqrt{2d}}\sum_{s'=1}^d \ket{0}\ket{s'}\ket{\tilde{\bm{\alpha}}_{s'}^{(t)}} + \ket{1}\ket{s'}\ket{\tilde{\bm{\alpha}}_{s'}^{(t)}}.
\end{equation}
Next, we use a CNOT gate to flip the sign of $\tilde{\bm{\alpha}}_{s'}$, where the control and target qubits are the first qubit and the specific qubit that records the sign information of $\bm{\alpha}_{s'}^{(t)}$, i.e., the state in Eqn.~(\ref{eqn:impt_a_state_prep_0-0}) transforms to  
\begin{equation}\label{eqn:impt_a_state_prep_0-1}
	\frac{1}{\sqrt{2d}}\sum_{s'=1}^d \left(\ket{0}\ket{s'}\ket{\tilde{\bm{\alpha}}_{s'}^{(t)}} + \ket{1}\ket{s'}\ket{-\tilde{\bm{\alpha}}_{s'}^{(t)}}\right).
\end{equation}
Absorbing the first qubit into the index register $\ket{s}$, we obtain 
 \begin{equation}\label{eqn:impt_a_state_prep_0}
 \frac{1}{\sqrt{2d}} \sum_{s=1}^{2d} \ket{s}\ket{\tilde{\bm{\alpha}}_{s}^{(t)}} ~.
 \end{equation}   
Notably, even though there exists an error $\varsigma$ with $|\tilde{\bm{\alpha}}_{s}^{(t)} -\bm{\alpha}_{s}^{(t)}|\leq \varsigma$ as discussed in Appendix \ref{appen:sec:state_prep}, the above implementation procedure guarantees 
\begin{equation}
	\tilde{\bm{\alpha}}_{s}^{(t)}=-\tilde{\bm{\alpha}}_{s+d}^{(t)}~,\forall s\in[d]~,
\end{equation}
since $\tilde{\bm{\alpha}}_{s+d}^{(t)}$ is produced by flipping the sign qubit of $\tilde{\bm{\alpha}}_{s+d}^{(t)}$. This property will be employed in the utility bound analysis. Once the state in Eqn.~(\ref{eqn:impt_a_state_prep_0}) is prepared, we apply a quantum adder, multiplier, and exponential operation   \cite{vedral1996quantum,haner2018optimizing} to obtain the state
\begin{align}\label{eqn:impt_a_state_prep_0-2}
& \frac{1}{\sqrt{2d}} \sum_{s=1}^{2d} \ket{s}\ket{\tilde{\bm{\alpha}}_{s}^{(t)}} \xrightarrow{+ 2L_1} \frac{1}{\sqrt{2d}} \sum_{s=1}^{2d} \ket{s}\ket{\tilde{\bm{\alpha}}_{s}^{(t)} + 2L_1 }\nonumber\\
& \xrightarrow{\times \frac{1}{\lambda}} \frac{-1}{\sqrt{2d}} \sum_{s=1}^{2d} \ket{s}\ket{-\frac{\tilde{\bm{\alpha}}_{s}^{(t)} + 2L_1}{\lambda} }  \nonumber\\
& 	\xrightarrow{\exp(\cdot)} \ket{\varphi(\bm{\alpha}^{(t)})} =  \frac{1}{\sqrt{2d}} \sum_{s=1}^{2d} \ket{s}\ket{\text{exp}\left(-\frac{\tilde{\bm{\alpha}}_{s}^{(t)} + 2L_1}{\lambda}\right) }~,
\end{align}
where we omit the absolute value operation since  $\tilde{\bm{\alpha}}_{s}^{(t)} + 2L_1 \geq 0$ for $\forall s\in [2d]$. 

Last, we introduce an ancillary qubit and apply a controlled rotation operation used in \cite{harrow2009quantum} to estimate the target state $\ket{\mathcal{W}(\bm{\alpha}^{(t)})}$, i.e.,
\begin{align}\label{eqn:impt_a_state_prep_0-3}
& \frac{1}{\sqrt{2d}} \sum_{s=1}^{2d} \ket{s}\ket{\text{exp}\left(-\frac{\tilde{\bm{\alpha}}_{s}^{(t)} + 2L_1}{\lambda}\right) }\ket{0} \xrightarrow[\text{on the 2cd quantum register}]	 {\text{conditional rotation}} \nonumber\\
 & 
\ket{\widetilde{\mathcal{W}}(\bm{\alpha}^{(t)})}=	\sum_{s=1}^{2d}\frac{1}{\sqrt{2d}}\ket{s}\ket{\tilde{\bm{\alpha}}^{(t)}_s}\nonumber\\
&
\times \left(\sqrt{ \frac{ e^{-\frac{|\tilde{\bm{\alpha}}^{(t)}_s+2L_1|}{\lambda}}}{C_1} }\ket{0}+ \sqrt{1 - \frac{e^{-\frac{|\tilde{\bm{\alpha}}^{(t)}_s+2L_1|}{\lambda}}}{C_1} }\ket{1}\right)~,
\end{align}   
where $C_1$ depends on $\lambda$ and $L_1$ as defined in Line 3-7 of Alg.~\ref{alg:Q_lasso_impt_pri_me}.	

Under the above description, the runtime cost to prepare the state in Eqn.~(\ref{eqn:impt_a_state_prep_0-0}) is $\tilde{O}(T^2\sqrt{N}/\varsigma)$, supported by Lemma \ref{thm:Q_state_prep}. The subsequent steps from Eqn.~(\ref{eqn:impt_a_state_prep_0-1}) to Eqn.~(\ref{eqn:impt_a_state_prep_0-2}), which aims to prepare the state $\ket{\varphi (\bm{\alpha}^{(t)})}$, only involve Hadamard transformations  and basic quantum arithmetic operations and can be completed in $O(\log (d))$ runtime. In the last step, the conditional rotation operation as formulated in Eqn.~(\ref{eqn:impt_a_state_prep_0-3}) can be achieved in $O(1)$ runtime \cite{harrow2009quantum}.  Overall, the runtime cost to prepare the state $\ket{\widetilde{\mathcal{W}}(\bm{\alpha}^{(t)})}$  is $\tilde{O}(T^2\sqrt{N}/\varsigma)$. 
\end{proof}

The result of Lemma 
\ref{lem:Q_state_prep_impt} and the optimization results of the FW algorithm allow us to quantify the utility and runtime cost of the QDP Lasso.
\begin{theorem}[Utility and runtime]\label{thm:utility_guarantee_impt}
Let $L_1\in \mathbb{R}_+$ be the Lipschitz constant of the loss $\mathcal{L}$ in Eqn.~(\ref{eqn:def_Lasso}), $0\leq \varsigma < L_1$ be the error threshold, and $C_f\in \mathbb{R}_+$ be the curvature constant of $\mathcal{L}$ in Definition \ref{def:curvature_const}. Given the dataset $D=\{\mathbf{X}\in\mathbb{R}^{N\times d}, \bm{y}\in\mathbb{R}^N\}$, with success probability $1-o(1)$, after $T=\frac{(N\epsilon)^{2/3}}{\ln^{1/3}(1/\delta)}$ iterations, ($\epsilon, \delta$)-QDP Lasso in Alg.~\ref{alg:Q_lasso_impt_pri_me} achieves the utility bound $
   R_{\mathcal{L}} \leq \tilde{O}\left(\frac{C_f^{1/3}}{N^{2/3}\epsilon^{2/3} }+ \varsigma + L_1 \right)$ in runtime  
 $\tilde{O}\left(\frac{C_f^{2}N^{5/2}\epsilon^{2}}{\varsigma}\right)$. 
\end{theorem}

 \begin{proof}[Proof of Theorem \ref{thm:utility_guarantee_impt}] We first analyze the Error   and utility bound of QDP and then derive its runtime complexity.
 
\textit{Error analysis and utility bound.} We now analyze how the selected basis $\hat{\bm{e}}_{K^{(t)}}$ as  described in Alg.~\ref{alg:Q_lasso_impt_pri_me} affects the error term $\vartheta$ given in Proposition \ref{prop:noisy_FW}, when the imperfection of the state preparation oracle $O_{\bm{\alpha}^{(t)}}$ is considered.  Specifically, supported by Lemma \ref{thm:Q_state_prep}, with success probability $1-2b$, the obtained result in Line 5 satisfies $|{\bm{\alpha}}^{(t)}_{k^{(t)}} - \tilde{\bm{\alpha}}_{k^{(t)}}^{(t)}|\leq \varsigma$ for $\forall k^{(t)}\in[2d]$. Consequently, the probability to accept $k^{(t)}$ as the output index shown in Line 6 is proportional to $\text{exp}(-|\tilde{\bm{\alpha}}_{k^{(t)}}^{(t)}+2L_1|/\lambda)$. Under this estimated acceptance rate, we now discuss the upper bound of  $\mathbb{E}\left[ \langle \hat{\bm{e}}_{K^{(t)}}, \nabla \mathcal{L}(\bm{\theta}^{(t)}) \rangle \right]$. Note that,  the symmetry property of $\bm{\alpha}^{(t)}=[\nabla \mathcal{L}(\bm{\theta}^{(t)}), -\nabla \mathcal{L}(\bm{\theta}^{(t)}) ]$ implies that its minimal entry, i.e., $\bm{\alpha}^{(t)}_{k^*}= \min_{\hat{\bm{e}}_s\in\mathcal{S}}  \langle \hat{\bm{e}}_s, \nabla \mathcal{L}(\bm{\theta}^{(t)}) \rangle$, is always less than zero.

Denote $\tilde{Z}= \sum_{s=1}^{2d}\exp\left(-\frac{\left| \tilde{\bm{\alpha}}^{(t)}_{s}+2 L_1 \right|}{\lambda} \right)$. The upper bound of $\mathbb{E}\left[ \langle \hat{\bm{e}}_{K^{(t)}}, \nabla \mathcal{L}(\bm{\theta}^{(t)}) \rangle \right]$ satisfies
\begingroup
\allowdisplaybreaks
 \begin{align}\label{eqn:thm_ut_pri_lasso_impt_2}
	& \mathbb{E}\left[ \langle \hat{\bm{e}}_{K^{(t)}}, \nabla \mathcal{L}(\bm{\theta}^{(t)}) \rangle \right] \nonumber\\
 	 = & \sum_{s=1}^{2d} \Pr(\text{Accept}|K^{(t)}=s)  \langle \hat{\bm{e}}_{s}, \nabla \mathcal{L}(\bm{\theta}^{(t)}) \rangle \nonumber\\
 	 = & \frac{\sum_{s=1}^{2d} \text{exp}\left(-\frac{\left| \tilde{\bm{\alpha}}^{(t)}_{s}+ 2L_1 \right|}{\lambda} \right)  \tilde{\bm{\alpha}}^{(t)}_{s} }{  \tilde{Z}} \nonumber\\
 	 = &  \tilde{\bm{\alpha}}^{(t)}_{k^*} - \tilde{\bm{\alpha}}^{(t)}_{k^*} +    \frac{\sum_{s=1}^{2d}  \text{exp}\left(-\frac{\left| \tilde{\bm{\alpha}}^{(t)}_{s}+2L_1 \right|}{\lambda } \right)  \tilde{\bm{\alpha}}^{(t)}_{s} }{\tilde{Z}}   \nonumber\\
 	\leq &  \bm{\alpha}^{(t)}_{k^*} + \varsigma  + L_1+ \varsigma +  \frac{\sum_{s=1}^{2d}  \exp\left(-\frac{\left| \tilde{\bm{\alpha}}^{(t)}_{s}+2L_1 \right|}{\lambda } \right)  \tilde{\bm{\alpha}}^{(t)}_{s}   }{\tilde{Z}} \nonumber\\
 	= &  \bm{\alpha}^{(t)}_{k^*} + 2\varsigma  +  L_1+  \frac{\sum_{s=1}^{2d}  \exp\left(-\frac{\left| \tilde{\bm{\alpha}}^{(t)}_{s}+2L_1 \right|}{\lambda } \right) \tilde{\bm{\alpha}}^{(t)}_{s}  }{\tilde{Z}} \nonumber\\
 	= & \bm{\alpha}^{(t)}_{k^*} + 2\varsigma +  L_1 + \frac{  \sum_{s^-\in \mathcal{S}^-}  \exp\left(-\frac{\left| \tilde{\bm{\alpha}}^{(t)}_{s^-}+2L_1 \right|}{\lambda } \right) \tilde{\bm{\alpha}}^{(t)}_{s^-}}{\tilde{Z}} \nonumber\\
 	& + \frac{\sum_{s^+\in\mathcal{S}^+}  \exp\left(-\frac{\left| \tilde{\bm{\alpha}}^{(t)}_{s^+}+2L_1 \right|}{\lambda } \right) \tilde{\bm{\alpha}}^{(t)}_{s^+}   }{\tilde{Z}}  \nonumber\\
 	\leq  &   \bm{\alpha}^{(t)}_{k^*} + 2 \varsigma + L_1+ \sum_{s^+\in \mathcal{S}^+ } \frac{ \exp\left(-\frac{\left| \tilde{\bm{\alpha}}^{(t)}_{s^+}+ 2L_1 \right|}{\lambda } \right)}{\tilde{Z}} \tilde{\bm{\alpha}}^{(t)}_{s^+} \nonumber\\
\leq & \bm{\alpha}^{(t)}_{k^*} + 2\varsigma + L_1+ \lambda~.
\end{align}
\endgroup 
 The first and second inequalities employ the facts that, for any $s\in[2d]$, $|\bm{\alpha}^{(t)}_s - \tilde{\bm{\alpha}}_s^{(t)}|\leq \varsigma$ as shown  in Lemma \ref{thm:Q_state_prep},  and  $|\bm{\alpha}^{(t)}_s|  \leq L_1$, since every entry of $\bm{\alpha}^{(t)}$ belongs to $\{\pm\nabla _s\mathcal{L}\}_{s=1}^d$ and $\|\nabla \mathcal{L} \|_2\leq L_1$. The third equality splits the summation into two groups, where the first and second group only includes the index that corresponds to the positive and negative  entries of the projected gradient, respectively, i.e, $\mathcal{S}^+= \{s^+|\tilde{\bm{\alpha}}^{(t)}_{s^+} >0\}$ and $\mathcal{S}^ - =\{s^-|\tilde{\bm{\alpha}}^{(t)}_{s^-} \leq 0\}$. The cardinality of these two groups are equal such that $|\mathcal{S}^+|=|\mathcal{S}^-|=d$. Then, in the third inequality, we exploits the following relation, i.e.,
 \begin{align}
 	\frac{\sum_{s^-\in\mathcal{S}^-}  \exp\left(-\frac{\left| \tilde{\bm{\alpha}}^{(t)}_{s^-}+ 2L_1 \right|}{\lambda } \right)   \tilde{\bm{\alpha}}^{(t)}_{s^-}  }{ \sum_{s=1}^{2d} \exp\left(-\frac{\left| \tilde{\bm{\alpha}}^{(t)}_{s} + 2L_1 \right|}{\lambda } \right)}
 \leq 0~,
 \end{align}
 since $\tilde{\bm{\alpha}}^{(t)}_{s^-}\leq 0$ and its coefficient is positive. The last  inequality is derived as follows, i.e., 
 \begingroup
 \allowdisplaybreaks
 \begin{align}
 	& \sum_{s^+\in\mathcal{S}^+} \frac{ \exp\left(-\frac{\left| \tilde{\bm{\alpha}}^{(t)}_{s^+} + 2L_1 \right|}{\lambda } \right)}{\sum_{s=1}^{2d}\exp\left(-\frac{\left| \tilde{\bm{\alpha}}^{(t)}_{s} + 2 L_1 \right|}{\lambda } \right)} \tilde{\bm{\alpha}}^{(t)}_{s^+} \nonumber	\\
 = & \frac{\sum_{s^+\in\mathcal{S}^+}  \exp\big(-\frac{\big| \tilde{\bm{\alpha}}^{(t)}_{s^+}+ 2L_1 \big|}{\lambda } \big)\tilde{\bm{\alpha}}^{(t)}_{s^+}}{\sum_{s^+\in\mathcal{S}^+}\exp\big(-\frac{\big| \tilde{\bm{\alpha}}^{(t)}_{s^+}+ 2L_1 \big|}{\lambda } \big) + \sum_{s^-\in\mathcal{S}^-}\exp\big(-\frac{\big| \tilde{\bm{\alpha}}^{(t)}_{s^-}+2L_1 \big|}{\lambda } \big) } \nonumber\\
 \leq & \sum_{s^+\in\mathcal{S}^+} \frac{ \exp\left(-\frac{\left| \tilde{\bm{\alpha}}^{(t)}_{s^+} + 2L_1 \right|}{\lambda } \right)}{\sum_{s^-\in\mathcal{S}^-}\exp\left(-\frac{\left| \tilde{\bm{\alpha}}^{(t)}_{s^-} + 2L_1 \right|}{\lambda } \right) }\tilde{\bm{\alpha}}^{(t)}_{s^+}  \nonumber\\
 = &   \sum_{s^+\in\mathcal{S}^+} \frac{ 1 }{\sum_{s^-\in\mathcal{S}^-}\exp\left(\frac{ \left| \tilde{\bm{\alpha}}^{(t)}_{s^+}+ 2L_1 \right| -  \left| \tilde{\bm{\alpha}}^{(t)}_{s^-}+ 2L_1 \right| }{\lambda } \right) }\tilde{\bm{\alpha}}^{(t)}_{s^+}  \nonumber\\
 \leq &\sum_{s^+\in\mathcal{S}^+}  \frac{  \tilde{\bm{\alpha}}^{(t)}_{s^+} }{\sum_{s^-\in\mathcal{S}^-}\frac{\tilde{\bm{\alpha}}^{(t)}_{s^+}}{\lambda} } \nonumber\\
 = & {\lambda}   ~,
 \end{align}
 \endgroup
 where the first inequality uses $\frac{1}{a + b}\leq \frac{1}{a}$ when $a , b >0$, and the last  inequality utilizes the facts $e^x\geq 1+ x$ and $L_1 \geq -\bm{\alpha}^{(t)}_{s^-}\geq 0$, which leads to $\text{exp} (\frac{ \left| \tilde{\bm{\alpha}}^{(t)}_{s^+}+ 2 L_1 \right| -  \left| \tilde{\bm{\alpha}}^{(t)}_{s^-}+ 2 L_1 \right| }{\lambda } ) \geq 1 + \frac{ \left| \tilde{\bm{\alpha}}^{(t)}_{s^+}+ 2 L_1 \right| -  \left| \tilde{\bm{\alpha}}^{(t)}_{s^-}+ 2 L_1 \right| }{\lambda } \geq \frac{  \tilde{\bm{\alpha}}^{(t)}_{s^+} + 2 L_1   -  ( \tilde{\bm{\alpha}}^{(t)}_{s^-} + 2 L_1 ) }{\lambda } \geq \frac{\tilde{\bm{\alpha}}^{(t)}_{s^+} }{\lambda } $.

 The result of  Eqn.~(\ref{eqn:thm_ut_pri_lasso_impt_2}) indicates that by choosing $\vartheta = \frac{4\varsigma}{ \mu_t C_f} + 2\frac{L_1+\lambda   }{\mu_t C_f}$, the output of Alg.~\ref{alg:Q_lasso_impt_pri_me} satisfies Eqn.~(\ref{eqn:prop_noisy_FW_0_1}). Therefore,  with success probability $1-o(1)$, Proposition \ref{prop:noisy_FW} yields
\begin{equation}\label{eqn:thm_ut_pri_lasso_impt_3}
R_{\mathcal{L}} \leq   \frac{2C_f}{T+2}(1+\frac{4\varsigma}{ \mu_t C_f} +  \frac{2L_1}{\mu_t C_f} + \frac{ 2\lambda   }{\mu_t C_f}) ~.
\end{equation}
By replacing the parameter $\lambda$ with its explicit form given in Line 6 of Alg.~\ref{alg:Q_lasso_impt_pri_me}, the utility bound follows
\begin{align}
	R_{\mathcal{L}} & \leq  \frac{2C_f}{T+2}(1+\frac{4\varsigma}{ \mu_t C_f} +  \frac{2L_1}{\mu_t C_f} + \frac{ 16\sqrt{2T\ln(1/\delta)}   }{\mu_t C_f \epsilon N}) \nonumber\\
	  & = \frac{2C_f}{T+2} + 4\varsigma + 2L_1 +  \frac{ 16\sqrt{2T\ln(1/\delta)}   }{ \epsilon N} ~.
\end{align}
The tight upper bound of $R_{\mathcal{L}}$ can be achieved by setting $T= \frac{C_f^{2/3}(N\epsilon)^{2/3}}{\ln^{1/3}(1/\delta)}$, i.e., 
\begin{equation}\label{eqn:thm_ut_pri_lasso_impt_4}
	R_{\mathcal{L}} \leq \tilde{O}\left(\frac{C_f^{1/3}}{N^{2/3}\epsilon^{2/3} }+ \varsigma + L_1 \right)~.
\end{equation}

\textit{The runtime complexity.}
We first quantify the required runtime cost at the $t$-th iteration, and then generalize to the entire $T$ iterations. Recall Alg.~\ref{alg:Q_lasso_impt_pri_me}. At the $t$-th iteration, the runtime cost is dominated by the implementation of the weighted sampling subroutine $\mathcal{W}(\bm{\alpha}^{(t)}))$ (Line 4). In particular, the state $\ket{\mathcal{W}(\bm{\alpha}^{(t)}))}$ for all $t\in[T]$  requires $\tilde{O}(T^2\sqrt{N}/\varsigma)$ runtime, as proved in Lemma \ref{lem:Q_state_prep_impt}. Given access to $\ket{\mathcal{W}(\bm{\alpha}^{(t)}))}$, when $L_1/\lambda \geq  \ln(1/\varsigma)$,  the probability     that we always see `1' during whole $M$ observations yields, 
\begin{align}
&	\left( 1 -  e^{-\frac{|\tilde{\bm{\alpha}}^{(t)}_s+2L_1|}{\lambda}}  \right)^M \nonumber\\
 < & \left( 1 -  e^{-\frac{4}{\ln(1/\varsigma)}}  \right)^M = (1 - \varsigma^4 )^M \leq e^{-4M\varsigma^4 } ~,
\end{align}

\begin{figure*}[htp]
\centering 
\includegraphics[width=0.895\textwidth]{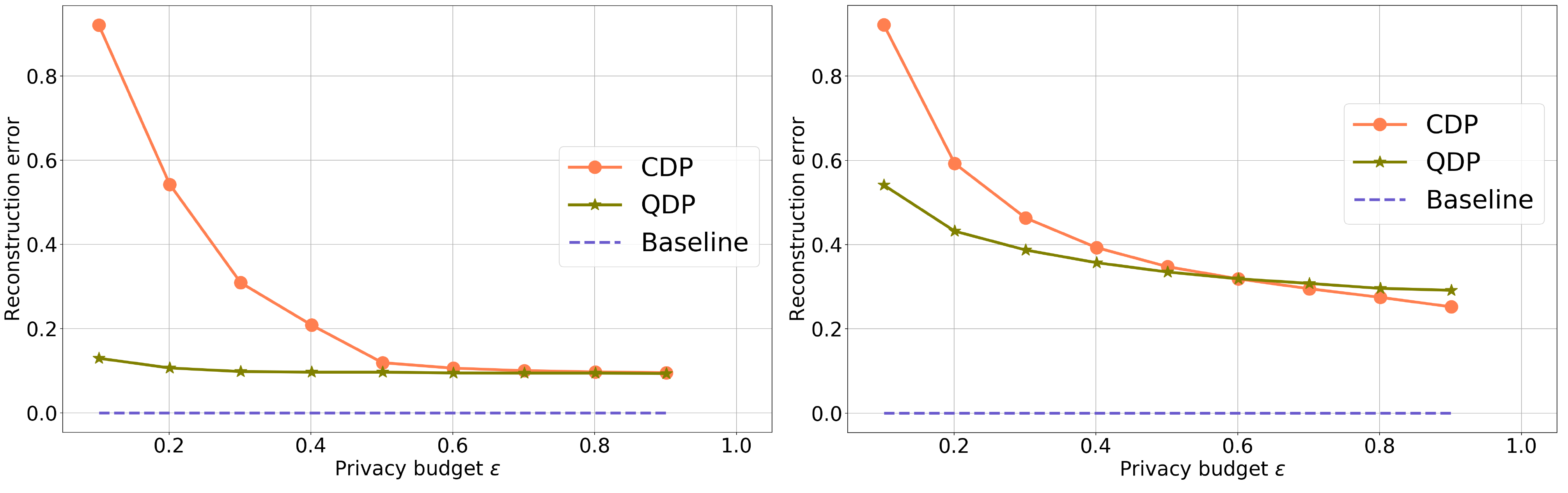} 
\caption{\small{The simulation results.  The label `Baseline' and `CDP' corresponds to the non-private Lasso in Alg.~\ref{alg:FW} and the proposal \cite{talwar2015nearly}. The left panel and the right panel separately illustrate the reconstruction error for the  datasets $D^{(1)}$ and $D^{(2)}$.} 
} 
\label{fig:private_lasso}
\end{figure*}   

where the first inequality employs $\tilde{\bm{\alpha}}^{(t)}_s+2L_1 < 4L_1$, and the second inequality employs $1-x\leq e^{-x}$. Let the rightest term in the above equation is upper bounded by $b$, we obtain
\begin{align}
	e^{-4M\varsigma^4} \leq b 
\Rightarrow  M \geq \frac{1 }{4\varsigma^4} \ln\frac{1}{b}~.
\end{align} 
This result indicates that, when $ M $ is larger than a certain constant, which is inversely proportional to $\varsigma$,  with probability $1-o(1)$, we can always observe the status `0' at least once. Consequently, the runtime complexity to accomplish the $t$-th iteration is $\tilde{O}(T^2\sqrt{N}/\varsigma)$. 

By leveraging the runtime to complete each iteration and the result of      Eqn.~(\ref{eqn:thm_ut_pri_lasso_impt_4}) that the optimal iterations follows $T= \frac{C_f^{2/3}(N\epsilon)^{2/3}}{\ln^{1/3}(1/\delta)}$, we conclude that the runtime complexity of Alg.~\ref{alg:Q_lasso_impt_pri_me} is 
\begin{align}
  \tilde{O}(T \cdot T^2\sqrt{N}/\varsigma) =	  \tilde{O}\left(\frac{C_f^{2}N^{5/2}\epsilon^{2}}{\varsigma}\right)~.
\end{align}
 \end{proof}
 
Theorem \ref{thm:utility_guarantee_impt}  allows us to infer all results presented in Table \ref{tab:my-table}.  Specifically, in conjunction with Lemma \ref{lem:lb_Lasso} and Theorem \ref{thm:utility_guarantee_impt}, an immediate observation is that the QDP Lasso is faster than the optimal classical non-private Lasso when $d>O(N^{5/2})$. Moreover, combining Lemma \ref{coro:Q_Lasso_lw} with Theorem \ref{thm:utility_guarantee_impt}, the QDP Lasso could be even faster than the optimal quantum non-private Lasso when $N>O(N^{5})$. These results separate quantum DP learning with its classical counterparts, where the implementation of classical private Lasso estimators generally require much more runtime than non-private cases. We note that in the quantum domain, the DP requirement separates the QDP Lasso with the quantum non-private Lasso to be two different learning problems. The former concerns finding a solution with the DP property and a good utility bound, while the latter    pursues the optimal solution in Eqn.~(\ref{eqn:def_Lasso}). These different aims enable that the proposed QDP Lasso may achieve a lower runtime complexity than that of the quantum non-private Lasso. Next, the utility bound of the QDP Lasso is near-optimal, since the study \cite{talwar2015nearly} has proven that the optimal utility of DP Lasso is ${\Omega}(1/(N\log N)^{2/3})$. Last, achieved near-optimal utility bound and runtime speedups indicate that the proposed QDP Lasso meets the two requirements of QDP learning.

 \begin{remark}\label{remark:impt_alg}
 \normalfont
We remark that the number of iterations of $T$ in Theorem \ref{thm:utility_guarantee_impt} is a positive integer. In other words, when $\epsilon$ is small and results in $T<1$, we should keep $T=1$. In this case, the runtime complexity of the proposed QDP Lasso estimator is $\tilde{O}(\sqrt{N}/\varsigma)$ and its  corresponding utility bound is $R_{\mathcal{L}}\leq \tilde{O}(2C_f+\varsigma+\frac{1}{\epsilon N})$ (see proof of Theorem \ref{thm:utility_guarantee_impt} for details). We next explain the examination procedure in Line 3-5 of Alg.~\ref{alg:Q_lasso_impt_pri_me} in the QDP Lasso. This step aims to ensure the legality of the privacy mechanism. Specifically,  when $L_1/\lambda \geq  \ln(1/\varsigma)$, the probability to sample the index $s$ for all $s\in[2d]$ will be near to zero, which renders the sampling process in Line 8 of Alg.~\ref{alg:Q_lasso_impt_pri_me}  to be ill-posed. Furthermore, when $L_1/\lambda \geq  \ln(1/\varsigma)$, the error term $\varsigma$ caused by the state preparation oracle will heavily affect the performance of the quantum private Lasso.
 	 
We further address the importance of examining the degree of the required privacy, driven by the distinguished runtime complexity between quantum private and non-private Lasso. Recall the conclusions of Theorem \ref{thm:runtime_q_lasso} and Theorem \ref{thm:utility_guarantee_impt}.  When the privacy budget $\epsilon$ is sufficiently small, the runtime cost for quantum private Lasso is $\tilde{O}(N^{5/2}\epsilon^2 /\varsigma)$, while a large privacy budget $\epsilon$ that is close to the non-private setting results in the $\tilde{O}(\sqrt{Nd})$ runtime. Without the examination in Line 3-5 of Alg.~\ref{alg:Q_lasso_impt_pri_me}, it is difficult to explain when the runtime of quantum Lasso should depend on the feature dimension $d$, since the privacy budget can be continuously degraded from the absolutely private case to the non-private case.

We last emphasize why the runtime of the quantum DP Lasso may be less than that of the quantum non-private Lasso when both of them employ the input/output quantum model formulated in Dentition \ref{def:input_model}. The central reason is that these two models focus on different aims, as stated in Introduction. Namely, quantum DP Lasso pursues both a lower runtime over its classical counterparts and a near-optimal utility bound for the specified differential privacy parameters, while quantum non-private machine learning algorithms only pursue low runtime when $\|\bm{\theta}^{(T)} - \bm{\theta}^*\|\leq \varsigma$. These different aims lead them to be two different learning problems. To elucidate how this separation allows QDP Lasso to advance quantum non-private Lasso, let us consider an extreme setting where the second requirement of QDP (i.e., attain a near-optimal utility bound) is discarded. In this scenario, the extreme QDP Lasso targets to output all possible solutions $\bm{\theta}$ with non-zero probability. This can be efficiently achieved by outputting the index $k^{(t)}$ randomly sampled from a uniform quantum state $\frac{1}{\sqrt{2d}}\sum_{s=1}^{2d}\ket{s}$ for $\forall t\in[T]$, which takes $\tilde{O}(1)$ runtime. Instead, the goal of quantum non-private Lasso is to find the optimal result $\bm{\theta}^*$ in Eqn.~(\ref{eqn:def_Lasso}). Therefore, according to Theorem \ref{thm:runtime_q_lasso}, for the same  $T$, the runtime complexity of this extreme QDP Lasso is lower than the quantum non-private Lasso, i.e., $\tilde{O}(T)$ versus $\tilde{O}(T^3\sqrt{Nd}/\varsigma)$. 	 
 \end{remark}

 \section{Numerical simulations}\label{sec:simulations}

In this section, we validate the performance of the QDP Lasso  using a synthetic dataset. We choose the reconstruction error, i.e., $\|\bm{\theta}^p-\bm{\theta}^*\|_2/\|\bm{\theta}\|_2$ that $\bm{\theta}^p$ is the output of the employed private Lasso estimator, as the metric to evaluate the  utility guarantee. To collect statistical information, all settings are repeated with $10$ times and then compute the average reconstruction error. The construction rule of the  synthetic dataset $\mathbf{X}\in\mathbb{R}^{n\times d}$ is as follows. Each entry of  $X$ is drawn i.i.d. from a uniform distribution $U(-1,1)$, and the $s^*$ non-zero entries of the optimal parameters $\bm{\theta}^*$ is drawn i.i.d. from $U(0,1)$.  Based on this construction rule, we  build  two synthetic datasets $D^{(1)}=(\mathbf{X}^{(1)}, \bm{y}^{(1)}) \in (\mathbb{R}^{400 \times 1000}, \mathbb{R}^{400})$ and $D^{(2)}=(\mathbf{X}^{(2)}, \bm{y}^{(2)})\in(\mathbb{R}^{1000 \times 5000},\mathbb{R}^{1000})$ with $s^*=10$ to evaluate performance of the QDP Lasso. The value of the  privacy budget $\epsilon$ is ranging from $0.1$ to $1$.

Figure \ref{fig:private_lasso} illustrates the simulation results. The reconstruction error of the classical DP Lasso with the Laplacian privacy mechanism \cite{talwar2015nearly} and the QDP Lasso is continuously decreasing with respect to the increased privacy budget $\epsilon$ for both $D^{(1)}$ and $D^{(2)}$. This empirical results accord with the theoretical results  as shown in  Theorem \ref{thm:utility_guarantee_impt}.    
  
\section{Applications}
A recent study indicates that quantum system noise enhances the threshold to attain quantum advantages \cite{babbush2020focus}. Concretely, \textit{quadratic speedups} will not enable quantum advantages on early generations of fault-tolerant devices and \textit{quartic speedups} look significantly more practical. This result denies the possibility to implement quantum kernel classifier \cite{li2019sublinear} or quantum kernel perceptron \cite{kapoor2016quantum} on  a modest fault-tolerant quantum computer to attain quantum advantages. Moreover, many quantum algorithms claiming exponential speedups are dequantized by quantum-inspired classical algorithms \cite{tang2019quantum}. The  proposed QDP Lasso, whose runtime is independent with feature dimension, paves a new way to attain advantages on early fault-tolerant quantum machines.  
  
\section{Conclusion}\label{sec:conclusion}
In this paper, we propose a QDP Lasso estimator to accomplish private sparse regression learning tasks. To the best of our knowledge, this is the first quantum private learning algorithm with runtime advantages and near-optimal utility. Moreover, we exhibit that the runtime of the QDP Lasso can be even lower than the optimal quantum non-private Lasso. The achieved results open an avenue to apply NISQ devices to attain runtime speedups. One of the interesting directions for the future study is to design other QDP algorithms that can provide both computational efficiency  and utility improvements.

\appendix

\section{Proof of Lemma~\ref{thm:Q_state_prep}}\label{appen:sec:state_prep}

The organization of the Appendix is as follows. Specifically, in Subsection \ref{append:subsec:imp_o_alpha}, we first demonstrate that the oracle  $O_{\tilde{\bm{\alpha}}^{(t)}}$, which prepares a quantum state that approximates the target state $O_{\bm{\alpha}^{(t)}}\ket{s}\ket{\bm{0}}=\ket{s}\ket{\bm{\alpha}^{(t)}_s}$ with an additive error, can be effectively implemented. We then analyze the required runtime complexity to implement $O_{\tilde{\bm{\alpha}}^{(t)}}$ in Subsection \ref{append:runt_imp_orac_o}. Last, we present the proof of Lemma \ref{lem:prep_OZ} and Lemma   \ref{lem:O_add}, which are employed to support the runtime analysis of implementing $O_{\tilde{\bm{\alpha}}^{(t)}}$, in Subsection \ref{subsec:lem_prep_OZ} and Subsection \ref{subsec:O_add},  respectively.

\subsection{Implementation of the state preparation oracle }  \label{append:subsec:imp_o_alpha}

The state preparation oracle $O_{\bm{\alpha}^{(t)}}$ in Alg.~\ref{alg:Q_lasso} aims to prepare the quantum state $O_{\bm{\alpha}^{(t)}}\ket{s}\ket{\bm{0}}=\ket{s}\ket{\bm{\alpha}^{(t)}_s}$, where
\begin{align}\label{eqn:sec_state_prep_1}
\bm{\alpha}_s^{(t)} 
 := &  \langle \hat{\bm{e}}_s,  \nabla L(\bm{\theta}^{(t)})\rangle \nonumber\\
 = & \nabla_s L(\bm{\theta}^{(t)}) \nonumber\\
 = & - \frac{1}{{N}} \sum_{i=1}^n \mathbf{X}_{is} \left(\bm{y}_i    -\langle  \mathbf{X}_i, \bm{\theta}^{(t)} \rangle\right),   
\end{align}
for $1< s \leq d$, and $\bm{\alpha}_s^{(t)} =-\bm{\alpha}_{s-d}^{(t)}$ for $d<s\leq 2d$. 

The above equation implies that the oracle $O_{{\bm{\alpha}}^{(t)}}$ can be constructed with an extra $O(1)$ runtime complexity if we have access to 
\begin{equation}\label{eqn:app_a_state_prep_0}
	O_{\hat{\bm{\alpha}}^{(t)}} \ket{s}\ket{\bf{0}}=  \ket{s}\ket{\bm{\alpha}_s^{(t)}}~\forall s\in[d]~.
\end{equation}
To be concrete, when the query $\ket{s}$ satisfies $d<s\leq 2d$, we first query $O_{\hat{\bm{\alpha}}^{(t)}}$ by the basis $\ket{s'}$ with $s'=s-d$ to obtain $\ket{s'}\ket{\bm{\alpha}_{s'}^{(t)}}$.  We then flip the qubit that records the sign information of $\bm{\alpha}_{s'}$ to obtain $\ket{s'}\ket{-\bm{\alpha}_{s'}^{(t)}}$.
Based on this fact,  the implementation of the oracle $O_{\tilde{\bm{\alpha}}^{(t)}}$ that approximates to $O_{\bm{\alpha}^{(t)}}$ amounts to preparing an oracle that approximates to $O_{\hat{\bm{\alpha}}^{(t)}}$. With a slight abuse of notation, in the following subsections, we use $O_{\tilde{\bm{\alpha}}^{(t)}}$ to specify the oracle that approximates to $O_{\hat{\bm{\alpha}}^{(t)}}$. For elucidating, Table \ref{tab:oracles-range-domain} summarizes the definitions and intuitions of quantum oracles explored in this section.

 \begin{table*}[h]
\caption{\small{\textbf{The summary of quantum oracles.} The intuition and definition of quantum oracles exploited in Algs.~1-5 are illustrated. The dimensionality of quantum state depends on the precision to store a real number. Here we suppose that all real numbers (i.e., $\mathbf{X}_{ij}$, $\bm{y}_i$, $\hat{\bm{\alpha}}^{(t)}_s$, $\bm{z}_i^{(t)}$, and $\bm{\theta}^{(t)}_j$) are stored by the single-precision floating-point format, which requests $32$ qubits.}}
\begin{adjustbox}{width=1.9\columnwidth,center}
\begin{tabular}{|l|l|l|}
\hline
Oracle                        & Intuition                                                                 & Definition                                                                                                                                                                                                                                  \\ \hline
$O_{\mathbf{X}}$              & Load input data $\bm{X}\in \mathbb{R}^{N\times d}$ into the quantum state & \begin{tabular}[c]{@{}l@{}}$O_{\mathbf{X}}(\mathcal{G})=\sqrt{|\mathcal{G}|}^{-1}\sum_{i,j\in\mathcal{G}}\ket{i,j}\ket{\mathbf{X}_{ij}} \in \mathbb{C}^{32Nd}$, \\ where $\mathcal{G}$ denotes a set of indexes to be queried.\end{tabular} \\ \hline
$O_{\bm{y}}$                  & Load input label $\bm{y}$ into the quantum state                          & \begin{tabular}[c]{@{}l@{}}$O_{\bm{y}}(\mathcal{G})=\sqrt{|\mathcal{G}|}^{-1}\sum_{i\in\mathcal{G}}\ket{i}\ket{\bm{y}_{i}} \in \mathbb{C}^{32N}$, \\ where $\mathcal{G}$ denotes a set of indexes to be queried.\end{tabular}               \\ \hline
$O_{\hat{\bm{\alpha}}^{(t)}}$ & Prepare the quantum state of $\hat{\bm{\alpha}}^{(t)}$ in Eqn. (13)       & $O_{\hat{\bm{\alpha}}^{(t)}}: \ket{s}\ket{\bf{0}} \rightarrow  \ket{s}\ket{\bm{\alpha}_s^{(t)}}\in \mathbb{C}^{64d}$                                                                                                                        \\ \hline
$O_{\bm{z}}$                  & A quantum subroutine to construct $O_{\hat{\bm{\alpha}}^{(t)}}$           & $O_{\bm{z}}:\frac{1}{\sqrt{N}}\sum_{i=1}^N\ket{i}\ket{\bm{0}} \rightarrow \frac{1}{\sqrt{N}}\sum_{i=1}^N\ket{i}\ket{\bm{z}_i^{(t)}} \in \mathbb{C}^{32N}$                                                                                   \\ \hline
$O_{\bm{\theta}^{(t)}}$       & Load trainable parameters into the quantum state                          & $O_{\bm{\theta}^{(t)}}:\ket{\bm{0}}\rightarrow \frac{1}{\sqrt{d}}\sum_{j=1}^d\ket{j}\ket{\bm{\theta}^{(t)}_j} \in \mathbb{C}^{32d}$                                                                                                             \\ \hline
\end{tabular}
\end{adjustbox}
\label{tab:oracles-range-domain}
\end{table*}

\subsubsection{Implementation of $O_{\hat{\bm{\alpha}}^{(t)}}$}

The implementation of the oracle $O_{\hat{\bm{\alpha}}^{(t)}}$, which leverages the reformulated expression of $\bm{\alpha}^{(t)}$ in Eqn.~(\ref{eqn:sec_state_prep_1}),   is summarized in Algorithm \ref{alg:QDP-StateP}. The detailed explanation of each step is as follows.  
 
\begin{algorithm}[h!]
   \caption{Quantum State Preparation oracle $O_{\hat{\bm{\alpha}}^{(t)}}$}
   \label{alg:QDP-StateP}
\begin{algorithmic}[1]
   \STATE {\bfseries Input:}   Input  oracles $O_{\mathbf{X}}$, $O_{\bm{y}}$, and the classical input $ \bm{\theta}^{(t)} $. 
  \STATE Initialize $\ket{s}_{r_0}\ket{\bf{0}}_{r_1}$ for quantum registers $r_0$ and $r_1$, and apply Hadamard transformation to the $r_1$ register  
  $$  \ket{s}_{r_0}\ket{\bf{0}}_{r_1}\xrightarrow[\text{transformation}]{\text{Hadamard}}\ket{s}_{r_0} \left(\frac{1}{\sqrt{N}} \sum_{i=1}^{N} \ket{i}_{r_1}\right); \label{lin:encode-2}
  $$
    \STATE Introduce the quantum register $r_2$ and apply oracle $O_{\bm{z}}$ formulated in Lemma \ref{lem:prep_OZ} to the register $r_1$ and $r_2$, i.e., let $\bm{z}_i^{(t)} = \bm{y}_i    -\langle  \mathbf{X}_i, \bm{\theta}^{(t)} \rangle$, 
    $$
    \frac{1}{\sqrt{N}} \ket{s}_{r_0}\sum_{i=1}^{N} \ket{i}_{r_1}\rightarrow\frac{1}{\sqrt{N}}\ket{s}_{r_0}\sum_{i=1}^{N} \ket{i}_{r_1}\ket{\bm{z}_i^{(t)}}_{r_2};
    $$\label{lin:encode-3}
  \STATE Introduce the quantum register $r_3$ and employ $O_{\mathbf{X}}$ to generate     
  $$ 
  \frac{1}{\sqrt{N}} \ket{s}_{r_0}\sum_{i=1}^{N} \ket{i}_{r_1}\ket{\bm{z}_i^{(t)}}_{r_2}\ket{\mathbf{X}_{is}}_{r_3};
  $$\label{lin:encode-4}
  \STATE Introduce the quantum register $r_4$, and apply the quantum operation   $U_{\text{inner}}$ as formulated in Lemma \ref{lem:O_add} to the quantum registers $r_2$ and $r_3$ to  obtain the state 
  $$
 \ket{s}_{r_0} \ket{\bm{\alpha}_s^{(t)}}_{r_4};
  $$\label{lin:encode-6}
     \STATE {\bfseries Output:} Output $\ket{s}\ket{\bm{\alpha}^{(t)}_s}$\label{lin:encode-7}; 
\end{algorithmic}
\end{algorithm}

The inputs of  Alg.~\ref{alg:QDP-StateP} are {two}  oracles, i.e., $O_{\mathbf{X}}$ and $O_{\bm{y}}$ in Definition \ref{def:input_model}, and the classical input $\bm{\theta}^{(t)}$  for $t\in [T]$. From Line 2 to Line 3 of Alg.~\ref{alg:QDP-StateP}, we employ Hadamard transformations and an oracle $O_{\bm{z}}$, i.e.,
\begin{equation}\label{eqn:O_z}
	O_{\bm{z}}:\frac{1}{\sqrt{N}}\sum_{i=1}^N\ket{i}_{r_1} \rightarrow \frac{1}{\sqrt{N}}\sum_{i=1}^N\ket{i}_{r_1}\ket{\bm{z}_i^{(t)}}_{r_2},
\end{equation}
 to load the vector $\bm{z}^{(t)}$, where $\bm{z}_i^{(t)}:=\bm{y}_i-\langle  \mathbf{X}_i, \bm{\theta}^{(t)}\rangle$ in Eqn.~(\ref{eqn:sec_state_prep_1}) for any $i\in[N]$, into the quantum register $r_2$. The construction of $O_{\bm{z}}$ requires $\tilde{O}(T^2)$ runtime, whose proof is given in Subsection \ref{subsec:lem_prep_OZ}.
\begin{lemma} \label{lem:prep_OZ}
Given access to  oracles $O_{\mathbf{X}}$ and $O_{\bm{y}}$ in Definition \ref{def:input_model}, and the classical input $\bm{\theta}^{(t)}$, the oracle $O_{\bm{z}}$ in Eqn.~(\ref{eqn:O_z}) can be implemented in $\tilde{O}(T^2)$ runtime. 
\end{lemma}

In Line 5 of Alg.~\ref{alg:QDP-StateP}, we aim to compute $\bm{z}_i^{(t)}\mathbf{X}_{is}$ and record the result in the quantum register $r_4$. Specifically, we apply the oracle $U_{\text{inner}}$ to the quantum registers $r_1$ and {$r_4$} to compute $\bm{\alpha}_s^{(t)} = \sum_i \bm{z}_i^{(t)}\mathbf{X}_{is}$. The implementation of $U_{\text{inner}}$ is summarized by the following lemma. 
\begin{lemma}\label{lem:O_add}
Given the access to the oracle $O_{\mathbf{X}}$ and  $O_{\bm{z}}$, there exists a  quantum operation $U_{\text{inner}}$ that estimates the inner product $\langle \mathbf{X}^{\top}_s, \bm{z}^{(t)} \rangle$, $\forall s\in[d]$ within the error threshold  $\varsigma$ and outputs the state formulated in Line 5  of Alg.~\ref{alg:QDP-StateP} with success probability $1-2b$. 
Suppose that the runtime to implement $O_{\bm{z}}$ is $T_{\bm{z}}$, the runtime complexity to implement $	U_{\text{inner}}$ is ${O}(T_{\bm{z}} \sqrt{N}\log(1/b)/\varsigma)$. 
\end{lemma}
The proof of Lemma \ref{lem:O_add} is given in Subsection \ref{subsec:O_add}.    

\subsection{Runtime complexity}\label{append:runt_imp_orac_o}

 \begin{proof}[Proof of Lemma \ref{thm:Q_state_prep}]
 The runtime complexity of the implementation of $O_{\hat{\bm{\alpha}}^{(t)}}$  as presented in Alg.~\ref{alg:QDP-StateP}  can be effectively obtained by combining the results of Lemma \ref{lem:prep_OZ}, \ref{lem:O_add}, and \ref{lem:prep-O-theta}. Note that, the error and uncertainty  is introduced by Line 5 of Alg.~\ref{alg:QDP-StateP}. Due to the result of Lemma \ref{lem:O_add}, the target state in Line 5 of  Alg.~\ref{alg:QDP-StateP} can only be approximately prepared in error $\varsigma$ with success probability $1-2b$. Therefore, the oracle constructed in Alg.~\ref{alg:QDP-StateP}  refers to $O_{\tilde{\bm{\alpha}}^{(t)}}$ instead of $O_{\bm{\alpha}^{(t)}}$, which prepares  the state $\ket{s}\ket{\tilde{\bm{\alpha}}_s^{(t)}}$ that satisfies $|{\bm{\alpha}}^{(t)}_s - \tilde{\bm{\alpha}}_s^{(t)}|\leq \varsigma $ for any $s\in[2d]$,  with success probability $1-2b$.

We now analyze the required runtime complexity Line by Line. The computation cost of Line 2  is $\tilde{O}(1)$, since only Hadamard transformation is employed. In Line 3, we use the oracle  $O_{\bm{z}}$ with the runtime complexity $\tilde{O}(T^2)$, supported by Lemma \ref{lem:prep_OZ}.  In Line 4, we call the oracle $O_{\bm{X}}$, which takes  $O(1)$ runtime. Therefore, the runtime complexity from Line 2 to Line 4 is $\tilde{O}(T^2)$. 

In Line 5, we employ $U_{\text{inner}}$ to prepare the target state. Following the conclusion of Lemma \ref{lem:prep_OZ}, the oracle $O_{\bm{z}}$ corresponds to the unitary transformations used in Line 2-4, which requires  $\tilde{O}(T^2)$  runtime complexity to implement it. Then, supported by  Lemma \ref{lem:O_add}, with success probability $1-2b$, the oracle $U_{\text{inner}}$ prepares the target state with error $\varsigma$ in runtime  $\tilde{O}(T^2\sqrt{N}/\varsigma)$. 

Overall, the runtime complexity of Alg.~\ref{alg:QDP-StateP}  is $\tilde{O}(T^2\sqrt{N}/\varsigma)$, which is dominated by Line 5. Since the oracle  $O_{\bm{\alpha}^{(t)}}$ can be efficiently implemented by using $O_{\hat{\bm{\alpha}}^{(t)}}$ with an extra $O(1)$ runtime as explained in Appendix \ref{append:subsec:imp_o_alpha}, the runtime complexity to implement the oracle $O_{\tilde{\bm{\alpha}}^{(t)}} \ket{s}\ket{\bf{0}}=  \ket{s}\ket{\tilde{\bm{\alpha}}_s^{(t)}}$ for $\forall s\in[2d]$ is also $\tilde{O}(T^2\sqrt{N}/\varsigma)$.

\end{proof}

\medskip
\subsection{Proof of Lemma \ref{lem:prep_OZ}}\label{subsec:lem_prep_OZ}

The proof of Lemma \ref{lem:prep_OZ}, or equivalently, the implementation of oracle $O_{\bm{z}}$, uses the following lemma.
\begin{lemma}\label{lem:prep-O-theta}
Denote the quantum oracle that prepares a state corresponding to $\bm{\theta}^{(t)}$ as $O_{\bm{\theta}^{(t)}}$, i.e.,
\begin{equation}\label{eqn:oracle_theta}
	O_{\bm{\theta}^{(t)}}:\ket{\bm{0}}\rightarrow \frac{1}{\sqrt{d}}\sum_{j=1}^d\ket{j}\ket{\bm{\theta}^{(t)}_j}~.
\end{equation}   
 Given the classical input $\bm{\theta}^{(t)}\in \mathbb{R}^d$ with $t\in[T]$, the oracle 	$O_{\bm{\theta}^{(t)}}$  formulated in Eqn.~(\ref{eqn:oracle_theta}) can be constructed in $\tilde{O}(T)$ runtime complexity. 
\end{lemma}  

\begin{proof}[Proof of Lemma \ref{lem:prep-O-theta}]
 Recall the updating rule of $\bm{\theta}^{(t)}$ described in Frank-Wolfe algorithm. The number of non-zero entries of $\bm{\theta}^{(t)}$ at the $t$ step is no greater than $t$. The sparsity of $\bm{\theta}^{(t)}$ implies that the oracle $O_{\bm{\theta}^{(t)}}$ can be efficiently implemented by using single-qubit and two-qubit gates with $\tilde{O}(T)$ complexity. The implementation of $O_{\bm{\theta}^{(t)}}$ is as follows, i.e.,
	\begin{align}
		& \ket{\bm{0}}\xrightarrow{O(\log d) \text{ $H$ gates}} \frac{1}{\sqrt{d}}\sum_{j=1}^d \ket{j}\ket{\bm{0}} \nonumber\\
		& \xrightarrow{\tilde{O}(T)\text{ multi-control gates}}\frac{1}{\sqrt{d}}\sum_{j=1}^d \ket{j}\ket{\bm{\theta}_j^{(t)}}~.
	\end{align}
	For each entry $\bm{\theta}_j^{(t)}$ with a constant precision, we need   ${O}(1)$  multi-control $(\log{d})$-qubit gates to encode it into the state $\ket{j}\ket{\bm{\theta}_j^{(t)}}$. Since the multi-control $(\log{d})$-qubit gate can be implemented by $O(poly(\log{d}))$ Toffoli gates and each Toffoli gate can be implemented by constant   single-qubit and two-qubit  gates \cite{barenco1995elementary}, the required number of single-qubit and two-qubit  gates  to implement  the multi-control $(\log{d})$-qubit gate is $O(poly(\log{d}))$. The sparsity of $\bm{\theta}^{(t)}$ implies that the total number of multi-control $(\log{d})$-qubit gates is $O(T)$. Alternatively, the required number of single-qubit and two-qubit gates to build $O_{\bm{\theta}^{(t)}}$ is $O(poly(\log{d})T)=\tilde{O}(T)$. 
\end{proof}

\begin{algorithm}[h]
  \caption{Quantum oracle $O_{\bm{z}}$ }
   \label{alg:Q-oracle-Oz}
\begin{algorithmic}[1]
   \STATE {\bfseries Input:}  Oracles $O_{\mathbf{X}}$, $O_{\bm{y}}$, and $O_{\bm{\theta}^{(t)}}$. 
   \STATE Define the set $J$ as the collection of indexes that corresponds to the non-zero entries of $\bm{\theta}^{(t)}$;	 
      \STATE  Prepare the uniform superposition state $\frac{1}{\sqrt{N}}\sum_{i=1}^N\ket{i};$
     \STATE Introduce  quantum registers $r_1$, $r_2$, and $r_3$, and separately  apply $O_{\bm{y}}$, $O_{\mathbf{X}}$, and $O_{\bm{\theta}^{(t)}}$ to obtain $$\frac{1}{\sqrt{N|J|}}\sum_{i=1}^N\ket{i}\ket{\bm{y}_i}_{r_1}\sum_{j\in J}\ket{j}\ket{\mathbf{X}_{ij}}_{r_2}\ket{\bm{\theta}^{(t)}_j}_{r_3};$$
    \STATE Apply quantum multiplier \cite{vedral1996quantum} to   $r_2$ and $r_3$,   store the result in quantum register $r_4$, and then  uncompute and delete $r_2$ and $r_3$, i.e.,  
    $$\frac{1}{\sqrt{N|J|}}\sum_{i=1}^N\ket{i}\ket{\bm{y}_i}_{r_1}\sum_{j\in J} \ket{j} \ket{\mathbf{X}_{ij}\bm{\theta}^{(t)}_j}_{r_4};$$ 
    \STATE {Apply the oracle $O_{\bm{v}}$} (see Eqn.~(\ref{eqn:O_z_1})) to the  register $r_4$   and record  $\bm{v}_i^{(t)} = \sum_j \mathbf{X}_{ij}\bm{\theta}^{(t)}_j$ in quantum register $r_5$, i.e., 
    $$ \frac{1}{\sqrt{N}|J|}\sum_{i=1}^N\ket{i}\ket{\bm{y}_i}_{r_1} \ket{\bm{v}_i^{(t)} }_{r_5}; $$
    \STATE{\bfseries Output:} Apply quantum subtractor \cite{vedral1996quantum} to record $\bm{z}_i^{(t)}$ in the quantum register $r_6$  conditionally controlled by $\ket{i}$, and then uncompute  and delete $r_1$ and $r_5$, i.e.,
    $  \frac{1}{\sqrt{N}}\sum_{i=1}^N\ket{i}\ket{\bm{z}_i^{(t)}}_{r_6}$. 
\end{algorithmic}
\end{algorithm}

We now employ the result of Lemma \ref{lem:prep-O-theta} to prove Lemma \ref{lem:prep_OZ}. 
\begin{proof}[Proof of Lemma \ref{lem:prep_OZ}]
	We illustrate the implementation of the oracle $O_{\bm{z}}$  in Alg.~\ref{alg:Q-oracle-Oz} and analyze its runtime complexity Line by Line. In particular, the runtime complexity of Line 3  is $\tilde{O}(1)$ by applying Hadamard transformation. The runtime complexity of Line 4  is $\tilde{O}(T)$, since the runtime complexity to implement $O_{\mathbf{X}}$ and $O_{\bm{y}}$ is $O(1)$ and the runtime complexity to implement $O_{\bm{\theta}^{(t)}}$ is  $\tilde{O}(T)$.  In Line 5 and Line 7,  the runtime complexity to conduct the multiplication and subtraction is $O(1)$, supported by \cite{vedral1996quantum}. 
	
In Line 6, the oracle  $O_{\bm{v}}$ is employed to compute $\bm{v}^{(t)}$ and record the result in the quantum register $r_5$. Note that, due to the sparsity of $\bm{\theta}^{(t)}$, the result of $\bm{v}_i^{(t)}$ only relates  to  non-zero entries of  $\bm{\theta}^{(t)}$. Motivated by such a fact, instead of encoding in total $d$ entries,  we only encode the entries whose indexes belong to the set $J$ into quantum states. 
The implementation of   $O_{\bm{v}}$ exploits the property that the cardinality $|J|$ is no greater than $T$ with $T\sim O(\log(N))$. In particular, given the quantum state formulated in Line 5 of Alg.~\ref{alg:Q-oracle-Oz}, the oracle $O_{\bm{v}}$ is composed of the following unitary transformations, i.e.,   
\begin{align}\label{eqn:O_z_1}
  &   \frac{1}{\sqrt{N|J|}}\sum_{i=1}^N\ket{i}\ket{\bm{y}_i} \sum_{j\in J} \ket{j} \ket{\mathbf{X}_{ij}\bm{\theta}^{(t)}_j}_{r_4} \nonumber\\
\rightarrow  & \frac{1}{\sqrt{N|J|}}\sum_{i=1}^N\ket{i}\ket{\bm{y}_i} \sum_{j\in J} \ket{j}  \ket{\mathbf{X}_{ij}\bm{\theta}^{(t)}_j}_{r_4}\ket{\bm{0}}_{q_2,...,q_{|J|}} \nonumber \\
\rightarrow   &  \frac{1}{\sqrt{N|J|}}\sum_{i=1}^N\ket{i}\ket{\bm{y}_i} \sum_{j\in J} \ket{j}  \ket{\mathbf{X}_{ij}\bm{\theta}^{(t)}_j}_{r_4}\ket{\mathbf{X}_{i,j+1}\bm{\theta}^{(t)}_{j+1}}_{q_2} \nonumber\\
& ~~~~~~~~~~~~~~~~~~~~~~~~~~~~ \ket{\cdots}_{q_3,...,q_{|J|-1}}\ket{\mathbf{X}_{i,j-1}\bm{\theta}^{(t)}_{j-1}}_{q_{|J|}} \nonumber\\
\rightarrow & \frac{1}{\sqrt{N|J|}}\sum_{i=1}^N\ket{i}\ket{\bm{y}_i} \sum_{j\in J} \ket{j}  \ket{\mathbf{X}_{ij}\bm{\theta}^{(t)}_j}_{r_4}\ket{\mathbf{X}_{i,j+1}\bm{\theta}^{(t)}_{j+1}}_{q_2} \nonumber\\
& ~~~~~~~~~~~~~~~~~~~~~~~~~~~~ \ket{\cdots}_{q_3,...,q_{|J|-1}}\ket{\mathbf{X}_{i,j-1}\bm{\theta}^{(t)}_{j-1}}_{q_{|J|}}\ket{\bm{v}_1}_{r_5}\nonumber\\
\rightarrow & \frac{1}{\sqrt{N|J|}}\sum_{i=1}^N\ket{i}\ket{\bm{y}_i}\ket{\bm{v}_i^{(t)}}_{r_5}~.
	\end{align}
The first arrow in Eqn.~(\ref{eqn:O_z_1}) is introducing {$|J|-1$} quantum registers. The second arrow indicates that,  the result $\mathbf{X}_{ij}\bm{\theta}_j^{(t)}$ for the different $j$ is recorded in $|J|-1$ quantum registers separately, by repeatedly calling  Line 4-5 of Alg.~\ref{alg:Q-oracle-Oz} to $\{q_2,...,q_{|J|}\}$ in total $|J|-1$ queries.   The last arrow shows that we uncompute and delete all quantum registers $\ket{j}$, $r_4$, and $\{q_i\}_{i=2}^{|J|}$. 
	
An observation of Eqn.~(\ref{eqn:O_z_1}) is that the runtime complexity to implement $O_{\bm{v}}$ is dominated by the second arrow.  Since  the runtime complexity of Line 4-5 is $\tilde{O}(T)$, the   runtime complexity to implement   $O_{\bm{v}}$ is $O((|J|-1)T)\leq \tilde{O}(T^2)$.

 Overall, the runtime complexity of Alg.~\ref{alg:Q-oracle-Oz} is $\tilde{O}({T^2})$, which is  dominated by implementing $O_{\bm{v}}$. 
\end{proof}

\subsection{Proof of Lemma  \ref{lem:O_add}}\label{subsec:O_add}
Lemma  \ref{lem:O_add} is a direct consequence of the following proposition \cite{kerenidis2019q}, which computes the inner product of two vectors. 
\begin{prop}[Modified from Lemma A.10, \cite{kerenidis2019q}]\label{prop:q_mean_alg}
 Suppose that we have access to two quantum oracles $O_{\mathbf{X}^{\top}}$ and $O_{\bm{z}}$ with $\mathbf{X}^{\top} \in\mathbb{R}^{d\times N}$ and $\bm{z}\in\mathbb{R}^{N}$,  
 \begin{eqnarray}
 O_{\mathbf{X}^{\top}}:\ket{i}\ket{\bm{0}}&\rightarrow& \frac{1}{\sqrt{N}}\ket{i}\sum_{j=1}^N\ket{j}\ket{\mathbf{X}^{\top}_{ij}} \\
  O_{\bm{z}}:\ket{\bm{0}}&\rightarrow&  \frac{1}{\sqrt{N}} \sum_{j=1}^N \ket{j}\ket{\bm{z}_j}  ~.
 \end{eqnarray}
 Denote the runtime complexity to implement $O_{\mathbf{X}^{\top}}$ and $O_{\bm{z}}$ is at most $O(T_{\max})$, and $\|\mathbf{X}_{i^*}\| = \max_i \|\mathbf{X}_{i}\|$,  there exists a quantum algorithm that with the probability at least $1-2b $, outputs the state 
 \begin{equation}
 	\ket{i}\ket{\overline{\langle \mathbf{X}^{\top}_i, \bm{z}\rangle}}~,
 \end{equation}
where $|\overline{\langle \mathbf{X}^{\top}_i, \bm{z}\rangle} -  \langle \mathbf{X}^{\top}_i, \bm{z}\rangle|\leq \varsigma$ in $O(\frac{T_{\max}\|\mathbf{X}_{i^*}\|\log(1/b)}{\varsigma})$ runtime.
\end{prop}

\begin{proof}[Proof of Lemma \ref{lem:O_add}]
 The implementation of $U_{\text{inner}}$ directly  employs the result of Proposition \ref{prop:q_mean_alg}. Recall that, given the state  
 \begin{equation}
 	\frac{1}{\sqrt{N}} \ket{s}\sum_{i=1}^{N} \ket{i}_{r_1}\ket{\bm{z}_i^{(t)}}_{r_2}\ket{\mathbf{X}_{is}}_{r_3}~,
 \end{equation}
 the operation $U_{\text{inner}}$ aims to estimate the inner production $\langle \mathbf{X}^{\top}_s, \bm{z}^{(t)}\rangle$ for any $s\in[2d]$. 
 
 Following the above observation, we apply Proposition \ref{prop:q_mean_alg} to the quantum register $r_2$ and $r_3$ to estimate $\langle \mathbf{X}^{\top}_s, \bm{z}^{(t)}\rangle$ in superposition. The result is stored in the quantum register $r_4$. We then uncompute the quantum registers $r_2$ and $r_3$, and output the obtained state.
 
With success probability $1-2b$, the quantum operation $U_{\text{inner}}$ prepares the estimated state $\frac{1}{\sqrt{2d}}\sum_{s} \ket{s} \ket{\bm{\alpha}_s^{(t)}}_{r_4}$ with error $\varsigma$ in runtime $${O}\left(\frac{T_{\bm{z}}\sqrt{N} \log(1/b)}{\varsigma}\right)~, $$ since $T_{\max}=T_{\bm{z}}$ and $\|\mathbf{X}_{i^*}\|\leq \sqrt{N}$ with the assumption $\|\mathbf{X}\|_{\infty}\leq 1$. 	 
\end{proof}

We remark that the core ingredients of Proposition \ref{prop:q_mean_alg} are the amplitude amplification and phase estimation \cite{brassard2002quantum,brassard2011optimal}. Due to a huge number of control qubits gates used in  quantum phase estimation, the computation-resource requirement may be unfriendly to the near-term quantum devices. Toward this issue, it is possible to attempt to employ a more advanced subroutine instead of the original one such as \cite{aaronson2020quantum}.

\ifCLASSOPTIONcaptionsoff
  \newpage
\fi

\begin{IEEEbiographynophoto}{Yuxuan Du} is currently a Senior Researcher at JD Explore Academy, and also a member of Doctor Management Trainee at JD. com. Prior to that, he received a Bachelor of Physics (elite class) from Sichuan University, and a Ph.D. degree of computer science from The University of Sydney. His research interests include fundamental algorithms for quantum machine learning, quantum learning theory, and quantum computing.  
\end{IEEEbiographynophoto}

\begin{IEEEbiographynophoto}{Min-Hsiu Hsieh} received his BS and MS in electrical engineering from National Taiwan University in 1999 and 2001, and PhD degree in electrical engineering from the University of Southern California, Los Angeles, in 2008. From 2008-2010, he was a Researcher at the ERATO-SORST Quantum Computation and Information Project, Japan Science and Technology Agency, Tokyo, Japan. From 2010-2012, he was a Postdoctoral Researcher at the Statistical Laboratory, the Centre for Mathematical Sciences, the University of Cambridge, UK. From 2012-2020, he was an Australian Research Council (ARC) Future Fellow and an Associate Professor at the Centre for Quantum Software and Information, Faculty of Engineering and Information Technology, University of Technology Sydney, Australia. He is now the director of Hon Hai (Foxconn) quantum computing center. His scientific interests include quantum information, quantum learning, and quantum computation.
\end{IEEEbiographynophoto}

\begin{IEEEbiographynophoto}{Tongliang Liu} is the Director of Sydney AI Centre at the University of Sydney. He is also heading the Trustworthy Machine Learning Laboratory. He is broadly interested in the fields of trustworthy machine learning and its interdisciplinary applications, with a particular emphasis on learning with noisy labels, adversarial learning, transfer learning, unsupervised learning, and statistical deep learning theory. He is a recipient of Discovery Early Career Researcher Award (DECRA) from Australian Research Council (ARC) and was named in the Early Achievers Leaderboard of Engineering and Computer Science by The Australian in 2020.
\end{IEEEbiographynophoto}

\begin{IEEEbiographynophoto}{Shan You} is currently a Senior Researcher at SenseTime, and also a post doc at Tsinghua University. Before that, he received a Bachelor of mathematics and applied mathematics (elite class) from Xi'an Jiaotong University, and a Ph.D. degree of computer science from Peking University. His research interests include fundamental algorithms for machine learning and computer vision, such as AutoML, representation learning, light detector and face analysis. He has published his research outcomes in many top tier conferences and transactions.
\end{IEEEbiographynophoto}

\begin{IEEEbiographynophoto}{Dacheng Tao}(F’15)  is an advisor and chief scientist of the digital science institute in the University of Sydney, and the Director of the JD Explore Academy and a Vice President of JD.com. He mainly applies statistics and mathematics to artificial intelligence and data science, and his research is detailed in one monograph and over 200 publications in prestigious journals and proceedings at leading conferences. He received the 2015 Australian Scopus-Eureka Prize, the 2018 IEEE ICDM Research Contributions Award, and the 2021 IEEE Computer Society McCluskey Technical Achievement Award. He is a fellow of the Australian Academy of Science, AAAS, ACM and IEEE.
\end{IEEEbiographynophoto}

\end{document}